\documentclass[iop]{emulateapj}
\usepackage[usenames, dvipsnames]{color}
\usepackage{chngcntr}
\usepackage[title]{appendix}
\usepackage{scalerel}
\usepackage{tikz}
\usetikzlibrary{svg.path}

\usepackage[bottom]{footmisc}

\definecolor{orcidlogocol}{HTML}{A6CE39}
\tikzset{
  orcidlogo/.pic={
    \fill[orcidlogocol] svg{M256,128c0,70.7-57.3,128-128,128C57.3,256,0,198.7,0,128C0,57.3,57.3,0,128,0C198.7,0,256,57.3,256,128z};
    \fill[white] svg{M86.3,186.2H70.9V79.1h15.4v48.4V186.2z}
                 svg{M108.9,79.1h41.6c39.6,0,57,28.3,57,53.6c0,27.5-21.5,53.6-56.8,53.6h-41.8V79.1z M124.3,172.4h24.5c34.9,0,42.9-26.5,42.9-39.7c0-21.5-13.7-39.7-43.7-39.7h-23.7V172.4z}
                 svg{M88.7,56.8c0,5.5-4.5,10.1-10.1,10.1c-5.6,0-10.1-4.6-10.1-10.1c0-5.6,4.5-10.1,10.1-10.1C84.2,46.7,88.7,51.3,88.7,56.8z};
  }
}

\newcommand\orcidicon[1]{\href{https://orcid.org/#1}{\mbox{\scalerel*{
\begin{tikzpicture}[yscale=-1,transform shape]
\pic{orcidlogo};
\end{tikzpicture}
}{|}}}}

\usepackage[breaklinks,colorlinks,urlcolor=blue,citecolor=blue,linkcolor=blue]{hyperref}

\usepackage[flushleft]{threeparttable}
\usepackage{ textcomp }
\usepackage{soul}
\usepackage{amsmath}
\usepackage{lineno}
\usepackage{rotating}
\usepackage{longtable}

\slugcomment{Draft Version Sep 15, 2021.}

\shorttitle{Improved Constraints For The IFMR}
\shortauthors{Barrientos \& Chanam\'e}

\begin{document}

\title{Improved Constraints on the Initial-to-Final Mass Relation of White Dwarfs using Wide Binaries}

\author{Manuel Barrientos \altaffilmark{1} \orcidicon{0000-0002-6153-9304}
Julio Chanam\'e \altaffilmark{1} \orcidicon{https://orcid.org/0000-0003-2481-4546} }

\affil{$^{1}$ Instituto de Astrof\'isica, Pontificia Universidad Cat\'olica de Chile, Av. Vicuna Mackenna 4860, 782-0436 Macul, Santiago, Chile; \href{mailto:mbarrien@astro.puc.cl}{mbarrien@astro.puc.cl} and \href{mailto:jchaname@astro.puc.cl}{jchaname@astro.puc.cl}}

\begin{abstract}
We present observational constraints for the initial-to-final mass relation (IFMR) derived from 11 white dwarfs (WDs) in wide binaries (WBs) that contain a turnoff/subgiant primary. Because the components of WBs are coeval to a good approximation, the age of the WD progenitor can be determined from the study of its wide companion.  However, previous works that used WBs to constrain the IFMR suffered from large uncertainties in the initial masses because their MS primaries are difficult to age-date with good precision. Our selection of WBs with slightly evolved primaries avoids this problem by restricting to a region of parameter space where isochrone ages are significantly easier to determine with precision. The WDs of two of our originally selected binaries were found to be close double degenerates, and are not used in the IFMR analysis. We obtained more precise constraints than existing ones in the mass range 1-2 M$_{\odot}$, corresponding to a previously poorly constrained region of the IFMR. Having introduced the use of turnoff/subgiant-WD binaries, the study of the IFMR is not limited anymore by the precision in initial mass, but now the pressure is on final mass, i.e., the mass of the WD today. Looking at the full dataset, our results would suggest a relatively large dispersion in the IFMR at low initial masses.  More precise determinations of the mass of the WD components of our targets are necessary for settling this question.

\end{abstract}

\renewcommand{\thefootnote}{\arabic{footnote}}
\section{Introduction}
\label{sec:1}

Stars with initial masses below $\sim$8 M$_{\odot}$ will one day evolve into white dwarfs (WDs), the evolutionary endpoint of over 97 \% of the stars in our Galaxy. The properties of WD progenitors during their main-sequence (MS) lifetimes are theoretically very well understood and observationally very well constrained. Similarly, once the WD is born, its subsequent evolution is straightforward and mainly governed by simple cooling by radiation from its surface. However, many aspects of the progenitor's evolution from the thermally pulsing asymptotic giant branch (TP-AGB) to its landing on the WD cooling sequence remain elusive. During the TP-AGB, because of the multiple pulses suffered by stars, their outer shells are expelled to the interstellar medium, shedding an important percentage of their mass \citep{weidemann2000}. Taking this into account, we do not have the ability to predict how much mass is lost during this process for a given progenitor mass. The mapping between the WD mass and its progenitor's mass is known as the initial-to-final mass relation (hereafter IFMR) of WDs. This relationship quantifies the mass lost by a star over its lifetime and therefore has implications on wide-ranging astronomical phenomena from the pathways that produce Type Ia supernovae to the future evolution of our Solar System \citep{williams2009}.

Different approaches have been used to analyze this relationship. On the one hand, there is the theoretical approach that uses stellar evolution models to predict the mass of the WD for a given initial mass \citep[e.g.,][]{dominguez1999,weiss2009,renedo2010,choi2016}. The IFMRs from these works differ from each other because of the different input physics used in their stellar evolution codes, for example, the theoretical mass-loss rates, convection treatment, third dredge-up efficiency, rotational mixing. In fact, stellar evolution models usually predict less massive WDs than what the observations suggest, likely due to the handling of AGB physics \citep{cummings2019}. On the other hand, there is the semi-empirical approach that requires the determination of the present masses of WDs and the initial masses of their progenitors. The former can be measured today via several observational techniques, but the latter is not directly measurable as the original star does not exist anymore. Therefore, the typical methodology is to obtain the total age of the WD; which is the sum of two lifetimes: that of its progenitor from the zero-age MS (ZAMS) up to the first thermal pulse of the AGB ($\tau_{\text{prog}}$) and its cooling time ($\tau_{\text{cool}}$), and then use stellar evolution models to trace back the mass of the WD progenitor.

Since stellar ages are most readily obtained for coeval groups of stars, WDs in open clusters (OCs) have often been used to constrain the IFMR \citep[e.g.,][]{kalirai2005,dobbie2006,kalirai2007,cummings2018,williams2018,marigo2020,canton2021}. The comparison of a WD's cooling time to its cluster's age provides the necessary information to infer the initial mass of the WD's progenitor. However, constraining the IFMR using OC's WDs is difficult. For accurate spectral determinations, only WDs in nearby OCs can place strong constraints. These stellar groups tend to be young enough that lower mass stars have not evolved off the MS, making this method most sensitive to the high-mass end of the IFMR. Moreover, the stars in OCs tend to be metal-rich, thus the constraints found using this method are limited because of the small metallicity coverage ([Fe/H]$\gtrsim$0).

Wide binaries (WBs) containing at least one WD provide an alternative method for calibrating the IFMR. We can define a WB as a system of two bodies with similar astrometric properties (positions, velocities, and distances), usually with orbital separations a $\gtrsim$ 100 AU \citep[e.g.,][]{c&g2004,andrews2017,elbadry2018}. The two components of a binary are expected to be coeval and co-chemical \citep[e.g.,][]{greenstein1986,kh2009,fran2021,andrews2019,ramirez2019}, and therefore these systems can be considered as the smallest possible examples of a star cluster; any property easily determined for one of the components (say its age, metallicity) can be safely assigned to the other \citep{chaname2007}. For these binaries, the coeval components are far enough apart that they can be assumed to have evolved in isolation (e.g., \citealt{silvestri2001}; but see \citealt{js1996} for potential caveats). Thus, the components of wide enough WBs have not been subjected to mass transfer during their lifetimes. In contrast to OCs, WBs provide a larger coverage in metallicity and age, allowing to study the IFMR with a group of stars more representative of the Galactic stellar populations.

WBs that include a MS star and a WD have been used to constrain the IFMR \citep{catalan2008a,zhao2012}. Although the spectroscopic analyses for the components of a WB are typically more accurate than in OCs (because the former are nearby), these constraints have bigger uncertainties in the initial mass determination because of the low precision obtaining ages for MS stars via theoretical isochrones, hence weakly constraining the relation. More recently, \cite{andrews2015} used wide double white dwarfs to study the IFMR, opening a new road for constraining convective overshoot and dredge-up processes in stellar evolution. This method provides robust constraints for initial masses between 2 and 4 M$_{\odot}$, but it does not give information about the effects of metallicity on the IFMR. In addition, \cite{elbadryb2018} used a CMD with photometric information and distances of field WDs in \textit{Gaia} second data release \citep[DR2;][]{gaiadr2} to constrain the IFMR. This relationship is in good agreement with previous constraints. However, no information about the progenitor's chemical composition can be derived.

In this paper, we aim to provide new constraints on the IFMR of WDs by taking advantage of WBs selected such that stellar age determinations for these systems are more reliable than previous works. For this, we selected pairs where the primary is an evolved turnoff (TO) or subgiant (SG) star and the secondary a WD. In these WBs, the total age of the system, and thus of the WD, is obtained from the fitting of theoretical isochrones to the position of the TO/SG primary in a Hertzsprung-Russell (HR) diagram. As shown by \cite{cr12}, isochrone ages for TO/SG primaries have typical precision of better than about 15-20\% with Hipparcos-class parallaxes. The expectation is that these new calibrators will help us to finally populate the low-mass end of IFMR (M$_{\text{i}}$ $<$ 2.5 M$_{\odot}$), with precise determinations of the initial masses in particular. 

We begin by describing the procedures and results of our search for WBs in Section~\ref{sec:2}. In Section~\ref{sec:3} and Section~\ref{sec:4}, we present the determination of stellar atmospheric parameters for both components and ages of the primaries using theoretical isochrones. In Section~\ref{sec:5}, we show and discuss the resulting constraints for the IFMR, finishing with our conclusions in Section~\ref{sec:6}.

\begin{figure}[t!]
	\includegraphics[width=0.47\textwidth]{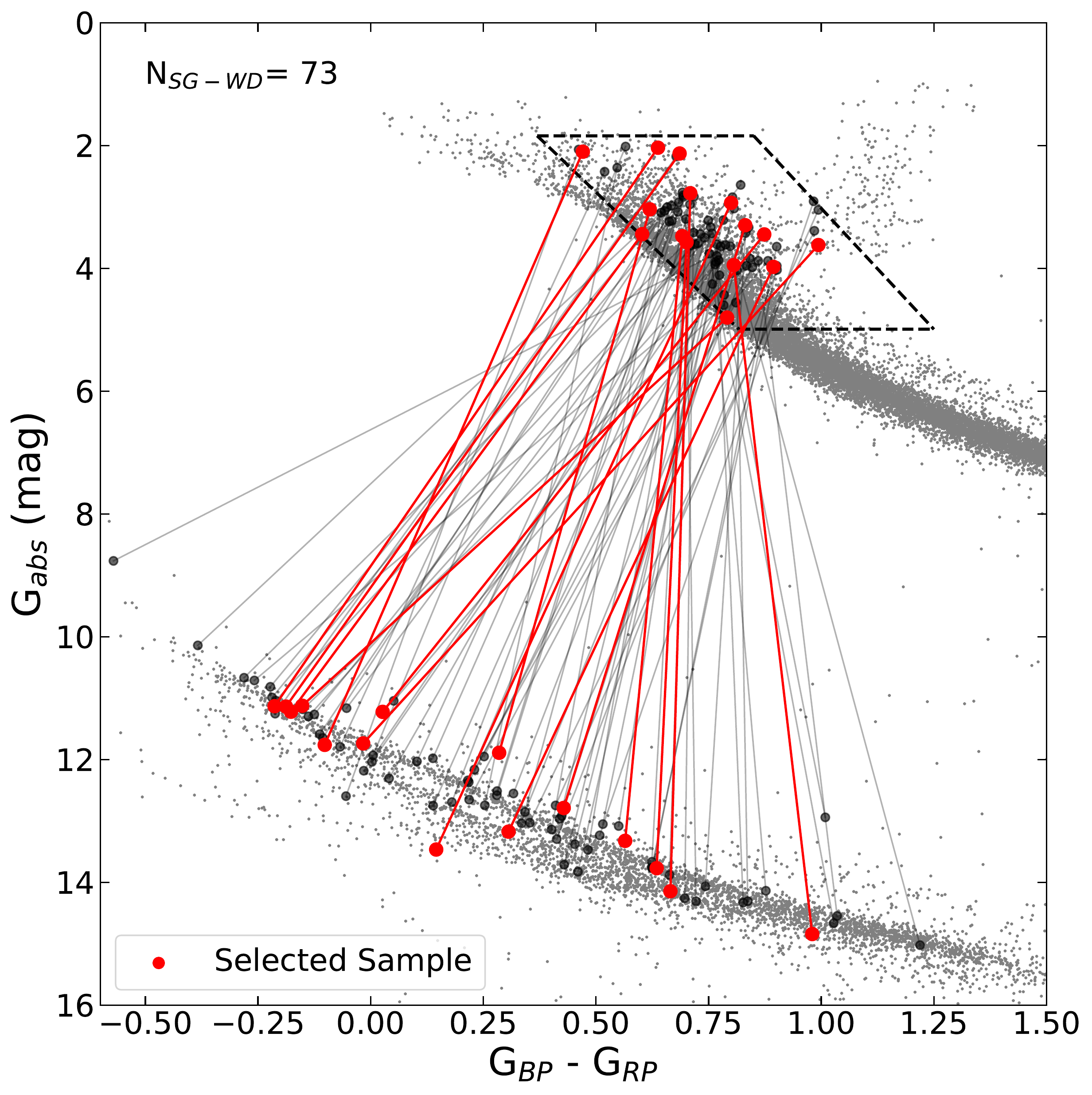}
    \caption{CMD showing the selected SG-WD binaries. The red linked dots show our selected wide binaries, the black linked dots are the 73 possible SG-WD pairs (19 from our search + 54 from \cite{elbadry2018}), grey dots in the background are stars from Gaia DR2 at distance $<$ 100 pc, and the black dotted lines show our selection region. }
    \vskip 3mm
   \label{fig:1}
\end{figure}

\section{Wide Binary search}
\label{sec:2}

We searched for WBs by matching positions, proper motions, and parallaxes from the Tycho-Gaia Astrometric Solution \citep[TGAS;][]{tgas2015} catalog, which contains the primaries, and the WD catalog by \cite{anguiano2017}, containing astrometric data for the secondary stars. TGAS is a subset of \textit{Gaia} data release 1 \citep[DR1;][]{gaiadr1} containing around 2 million stars with a full astrometric solution. The WD catalog by \cite{anguiano2017} is a subsample of 20247 hydrogen-rich (DA) WDs in SDSS data release 12 \citep[DR12;][]{kepler2016}. This catalog has positions and proper motions obtained by combining SDSS and recalibrated USNO-B astrometry \citep{munn2014}, SDSS photometric distances, and spectroscopic physical parameters such as effective temperatures, surface gravities, masses, and cooling times. 

We cannot use radial velocities as a criterion for identifying our WBs. For the WDs, it is not possible to measure this parameter directly given the gravitational redshift effect. The latter produces a wavelength shift in the absorption lines of the WD spectra proportional to its mass-radius ratio. To study this effect with precision, high-resolution spectra are needed in order to resolve the core of the Balmer lines \citep[e.g.,][]{silvestri2001,falcon2010,napi2020}; which it is not the case in this work.

For our search, we first updated the astrometry of both catalogs to the most recent and precise measurements from Gaia Early Data Release 3 \citep[EDR3;][]{gaiaedr3}.We followed a similar strategy as \cite{andrews2017}, \cite{gr2018}, and \cite{elbadry2018} (hereafter EB18). In order to avoid extra calculations and speed up the cross-matching process, we need to select those stars with proper motions ($\mu_{\alpha}$, $\mu_{\delta}$), and parallaxes ($\varpi$) well measured. For this, we created two dimensionless parameters to evaluate the quality of the data and require that
\begin{equation}
\small
    \frac{\mu}{\sigma_{\mu}} = \frac{\vec{|\mu|}}{\vec{|\sigma_{\mu}|}}  = \frac{\sqrt{\mu_{\alpha}^2 + \mu_{\delta}^2}}{\sqrt{\sigma_{\mu_{\alpha}}^2 + \sigma_{\mu_{\delta}}^2}} \geq 3 \qquad \textrm{  and  }\qquad \frac{\varpi}{\sigma_{\varpi}} \geq 3 \textrm{ ,}
  \label{eq:1}    
\end{equation}
where $\mu_{\alpha}$ (accounting for the $\cos(\delta)$ factor) and $\mu_{\delta}$ are the proper motions in right ascension ($\alpha$) and declination ($\delta$), respectively. The uncertainties for this measurements are given by $\sigma_{\mu_{\alpha}}$ and $\sigma_{\mu_{\delta}}$, and the parallax and its error are $\varpi$ and $\sigma_{\varpi}$.
 
The first selection parameter is related to the projected separation of our WBs defined as s = 1000 au $\times (\Delta \theta/\text{arcsec}) \times (\varpi/\text{mas})^{-1}$, where $\Delta \theta$ is the angular separation between stars (A and B) in the sky
\begin{equation}
\small
    \Delta \theta \simeq \sqrt{(\alpha_A-\alpha_B)^2  \cos{\delta_A}\cos{\delta_B} + (\delta_A - \delta_B)^2} \textrm{ .}\\
\label{eq:2}
\end{equation}

The limit in projected separation for a genuine WB has been discussed in the literature \citep[e.g.,][]{YCG2004,jt2010,andrews2017}. \cite{andrews2017,andrews2018} found that common proper motion, common parallax pairs selected from TGAS with projected separations s $>$ 4 $\times$ $10^4$ au are mostly composed of random, chance alignments. Nevertheless, genuine pairs beyond this limit do exist \citep{c&g2004,quinn2010}. Taking this into account, we allowed our search to projected separations
\begin{equation}
    s \lesssim 5 \times 10^{4} \text{ au.}
    \label{eq:3}
\end{equation}

To identify the WBs we are interested in, we need both stars to be at the same distance from us. The stars in this sample have distances to the Sun $<$ 350 pc. In this regime, we checked that the differences between the 1/$\varpi$ approximation and Bayesian distances by \cite{bj2018} were within the errors of the measurements; which was the case. We require that 
\begin{equation}
\small
    \Delta \varpi \leq 3\sigma_{\varpi_{\text{AB}}} \textrm{ ,}\\
    \label{eq:4}
\end{equation}
 where the $\sigma_{\varpi_{\text{AB}}}$ considered here is a combinations in quadrature of both errors $\sigma_{\varpi_{\text{A}}}$ and $\sigma_{\varpi_{\text{B}}}$.
 
We need both stars to move in the same direction at a similar speed. At the same time, we want to exclude unbound visual binaries and unresolved higher-order systems. To ensure this, we used the difference in transverse velocity $\Delta V_{\perp}$ = 4.74 $\times (\Delta \mu/\text{mas yr}^{-1}) \times (\varpi_A/\text{mas})^{-1}$ km s$^{-1}$, where $\varpi_A$ is the parallax of the brighter component, $\Delta \mu$ and its error $\sigma_{\Delta \mu}$ are given by
\begin{equation}
\small
    \Delta \mu = \sqrt{(\mu_{\alpha,A}-\mu_{\alpha,B})^2 + (\mu_{\delta,A}-\mu_{\delta,B})^2 } \textrm{ ,}\\
    \label{eq:5}
\end{equation}

\begin{equation}
\small
    \sigma_{\Delta \mu} = \frac{1}{\Delta \mu}\sqrt{(\sigma_{\mu_{\alpha,A}}^2 + \sigma_{\mu_{\alpha,B}}^2)\Delta \mu_{\alpha}^2 + (\sigma_{\mu_{\delta,A}}^2 + \sigma_{\mu_{\delta,B}}^2)\Delta \mu_{\delta}^2 } \textrm{ ,}\\
    \label{eq:6}
\end{equation}
and require that
\begin{equation}
\small
    \Delta V_{\perp} < \Delta V_{\text{orb}} + 2 \sigma_{\Delta V_{\perp}} \textrm{,}\\
    \label{eq:7}
\end{equation}
with the projected physical velocity difference $\Delta V_{\text{orb}}$ and $\sigma_{\Delta V_{\perp}}$ defined as
\begin{equation}
\small
   \Delta V_{\text{orb}} = 2.1 \times \left(\frac{s}{1000 \text{ au}}\right)^{-1/2} \textrm{ km s$^{-1}$.}\\
   \label{eq:7}
\end{equation}

\begin{equation}
\small
   \sigma_{\Delta V_{\perp}} = 4.74 \times \sqrt{\frac{(\Delta \mu)^2}{\varpi_A^4}\sigma^2_{\varpi_A} + \frac{\sigma^2_{\Delta \mu}}{\varpi_A^2}} \textrm{ km s$^{-1}$.}\\
   \label{eq:7}
\end{equation}
The former quantity is directly associated with the maximum proper motion difference expected for a circular orbit of total mass of 5 M$_\odot$ \citep{elbadry2021}.

\begin{figure}[t!]
\centering
	\includegraphics[width=0.47\textwidth]{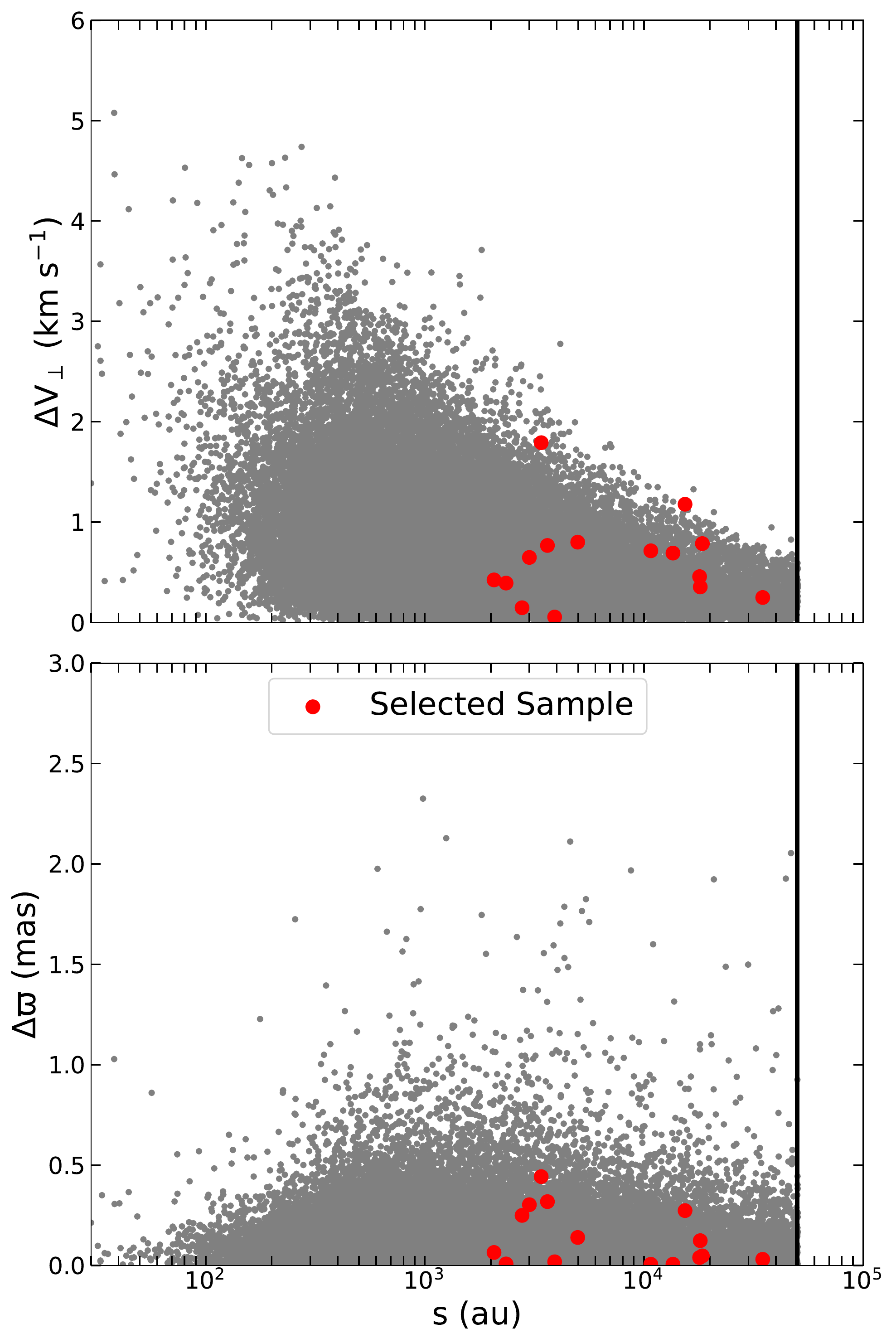}
    \caption{Difference in transverse velocity ($\Delta V_{\perp}$) and parallax ($\Delta \varpi$) for our wide binary sample in terms of projected separation (s). The red dots are the selected sample in this work, the black line shows the projected separation limit, and the grey background is the wide binary catalog by \cite{elbadry2018}.}
        \vskip 3mm
   \label{fig:2}
\end{figure}

\begin{table*}[t!]
\centering
\def\arraystretch{1.57}
\setlength{\tabcolsep}{5.2pt}
\caption{Wide Binaries in Our Sample: Turnoff/Subgiant Primaries and White Dwarf Secondaries}
\label{tab:1}
\begin{tabular}{c c c c c c c c c c}
\hline\hline
Pair N\textdegree & Name & Gaia source ID &  $\mu_{\alpha}$ & $\mu_{\beta}$ & $\varpi$ & G & $\Delta \theta$ & $\Delta V_{\perp}$ & s \\
 &  & (DR2) & (mas yr$^{-1}$) & (mas yr$^{-1}$) & (mas)  & (mag) & (arcsec) & (km s$^{-1}$) & (au)\\
\hline
\hline
\multicolumn{9}{c}{WBs our search}\\
\hline
\hline
1 & SG1 & 2751887496187636096 & -2.49 $\pm$ 0.03 & -8.76 $\pm$ 0.02 & 3.60 $\pm$ 0.02 & 9.35 & 38.73 & 0.72 & 10750.07\\ 
 & WD1 & 2751887427468160000 & -1.95 $\pm$ 0.21 & -8.81 $\pm$ 0.16 & 3.61 $\pm$ 0.20 & 18.44 & & & \\ \hline 
2  & SG2 & 593915842991753088 & -7.79 $\pm$ 0.02 & 4.00 $\pm$ 0.02 & 2.91 $\pm$ 0.02 & 11.23 & 52.15 & 0.46 & 17945.20\\ 
 & WD2 &  593918802224747776 & -8.02 $\pm$ 0.26 & 4.16 $\pm$ 0.20 & 2.95 $\pm$ 0.25 & 18.97 & & & \\ \hline 
3 & SG3 & 668561279381974400 & -35.96 $\pm$ 0.02 & -32.12 $\pm$ 0.01 & 6.91 $\pm$ 0.02 & 8.94 & 127.63 & 0.79 & 18478.09\\
 & WD3 & 668562516332550784 & -36.04 $\pm$ 0.19 & -30.98 $\pm$ 0.11 & 6.86 $\pm$ 0.17 & 17.81 & & & \\ \hline 
4 & SG4 & 3244617448738051200 & 35.90 $\pm$ 0.03 & 15.01 $\pm$ 0.02 & 4.69 $\pm$ 0.02 & 11.54 & 18.35 & 0.06 & 3911.98\\ 
 & WD4 & 3244617478801862656 & 35.86 $\pm$ 0.15 & 15.04 $\pm$ 0.12 & 4.67 $\pm$ 0.14 & 17.88 & & & \\ \hline 
5 & SG5 & 3637279439295576320 & -30.11 $\pm$ 0.02 & 0.68 $\pm$ 0.01 & 5.31 $\pm$ 0.02 & 8.58 & 95.92 & 0.36 & 18070.93\\ 
 & WD5 & 3637279538079751168 & -30.50 $\pm$ 0.17 & 0.62 $\pm$ 0.09 & 5.43 $\pm$ 0.13 & 18.19 & & & \\ \hline 
6 & SG6 & 3803142829929703552 & -17.55 $\pm$ 0.02 & -10.88 $\pm$ 0.02 & 5.19 $\pm$ 0.02 & 9.97 & 14.41 & 0.15 & 2778.05\\ 
 & WD6 & 3803142859993965952 & -17.60 $\pm$ 0.13 & -10.73 $\pm$ 0.11 & 4.94 $\pm$ 0.12 & 17.86 & & & \\ \hline 
7 & SG7 & 3975115244607118848 & 10.06 $\pm$ 0.02 & -38.11 $\pm$ 0.02 & 4.06 $\pm$ 0.02 & 10.68 & 14.72 & 0.77 & 3626.72\\ 
 & WD7 & 3975115240311668096 & 9.40 $\pm$ 0.27 & -38.17 $\pm$ 0.22 & 4.38 $\pm$ 0.32 & 18.63 & & & \\ 
\hline
\hline
\multicolumn{9}{c}{WBs from \cite{elbadry2018}*}\\
\hline
\hline
8 & SG8 & 93238895273066752 & -47.64 $\pm$ 0.03 & -6.74 $\pm$ 0.03 & 9.52 $\pm$ 0.04 & 8.78 & 32.33 & 1.79 & 3397.36\\
 & WD8 & 93238895273124480 & -49.32 $\pm$ 0.26 & -9.92 $\pm$ 0.27 & 9.96 $\pm$ 0.26 & 18.88 & & & \\ \hline 
9 & SG9 & 2198430859305721344 & 236.55 $\pm$ 0.07 & 129.06 $\pm$ 0.06 & 26.50 $\pm$ 0.07 & 5.11 & 54.82 & 0.43 & 2068.82\\ 
 & WD9 & 2198431172852758656 & 234.18 $\pm$ 0.02 & 128.95 $\pm$ 0.02 & 26.43 $\pm$ 0.02 & 14.13 & & & \\ \hline 
10 & SG10 & 6159300796901641856 & -10.88 $\pm$ 0.02 & 8.87 $\pm$ 0.01 & 7.15 $\pm$ 0.02 & 9.30 & 21.45 & 0.65 & 2999.69\\ 
 & WD10 & 6159300693821280000 & -10.18 $\pm$ 0.30 & 8.18 $\pm$ 0.20 & 6.85 $\pm$ 0.33 & 19.25 & & & \\ \hline 
11 & SG11 & 2879667068210826752 & -123.43 $\pm$ 0.02 & -90.42 $\pm$ 0.01 & 19.76 $\pm$ 0.02 & 6.40 & 46.40 & 0.39 & 2347.99\\ 
 & WD11 & 2879667033851088896 & -123.32 $\pm$ 0.10 & -88.78 $\pm$ 0.08 & 19.76 $\pm$ 0.11 & 17.77 & & & \\ \hline 
12 & SG12 & 64009276298551424 & 70.20 $\pm$ 0.14 & -69.82 $\pm$ 0.10 & 8.90 $\pm$ 0.13 & 9.33 & 137.15 & 1.18 & 15402.34\\ 
 & WD12 & 64009654256059008 & 70.02 $\pm$ 0.20 & -72.03 $\pm$ 0.14 & 9.18 $\pm$ 0.17 & 18.46 & & & \\ \hline 
13 & SG13 & 129073632786900480 & 30.31 $\pm$ 0.02 & -56.25 $\pm$ 0.02 & 5.79 $\pm$ 0.02 & 9.58 & 201.72 & 0.25 & 34839.76\\
 & WD13 & 129072494619153536 & 30.01 $\pm$ 0.29 & -56.19 $\pm$ 0.31 & 5.76 $\pm$ 0.27 & 19.09 & & & \\ \hline 
14 & SG14 & 1999563943535301760 & 193.59 $\pm$ 0.02 & 282.40 $\pm$ 0.02 & 20.20 $\pm$ 0.02 & 7.52 & 100.68 & 0.80 & 4984.47\\ 
 & WD14 & 1999564768169857792 & 195.59 $\pm$ 0.12 & 285.17 $\pm$ 0.14 & 20.34 $\pm$ 0.13 & 18.46 & & & \\ \hline 
15 & SG15 &  2153552647245580544 & -26.61 $\pm$ 0.01 & -37.63 $\pm$ 0.01 & 5.59 $\pm$ 0.01 & 9.29 & 75.87 & 0.69 & 13563.35\\ 
 & WD15 & 2153552814748001792 & -27.43 $\pm$ 0.38 & -37.69 $\pm$ 0.40 & 5.60 $\pm$ 0.33 & 19.82 & & & \\ \hline 
\hline\hline
\end{tabular}
\begin{tablenotes}
\small
\item \textbf{Notes.} *: The astrometric parameters of the pairs by ER18 were updated to \textit{Gaia} EDR3.
\end{tablenotes}
\smallskip
\end{table*}

Using this set of criteria, we found 37 WB candidates whose primaries are MS, TO, SG, and Giant stars (full sample in Table~\ref{tab:A.1}). The \textit{Gaia} magnitudes for the primary components are in the range 8.35 $<$ G $<$ 12.16 mag, and their WD secondaries lie between 16.84 $<$ G $<$ 20.33 mag. For the purposes of this paper, the pairs of interest are those where the primary star is a TO or SG star.

To select this type of primaries, we follow a similar procedure as in \cite{cr12} and use a CMD, where we included stars from \textit{Gaia} DR2 to use as a background for identifying the MS, SG, and Giant branch and a selection region of TO/SG stars as can be seen in Figure~\ref{fig:1} (black dotted box). By doing this, we can have approximate previous knowledge of the evolutionary stage of stars, but it does not guarantee the complete exclusion of MS stars or Giant stars.  

Using this last criterion, our sample was reduced to 19 SG-WD pairs. Of these, 13 have declinations below +20\textdegree, i.e., reachable from Southern hemisphere telescopes. We obtained spectra from 7 of these evolved primaries, the data is described in Section~\ref{sec:4.1} ahead.

\subsection{Wide Binaries in the Literature}
\label{sec:2.1}

To extend our sample of WBs, we searched for SG-WD for which stellar parameters could be found in the literature. In particular, we explored the WB catalog by EB18.

\begin{figure*}[t!]
\centering
	\includegraphics[width=\textwidth]{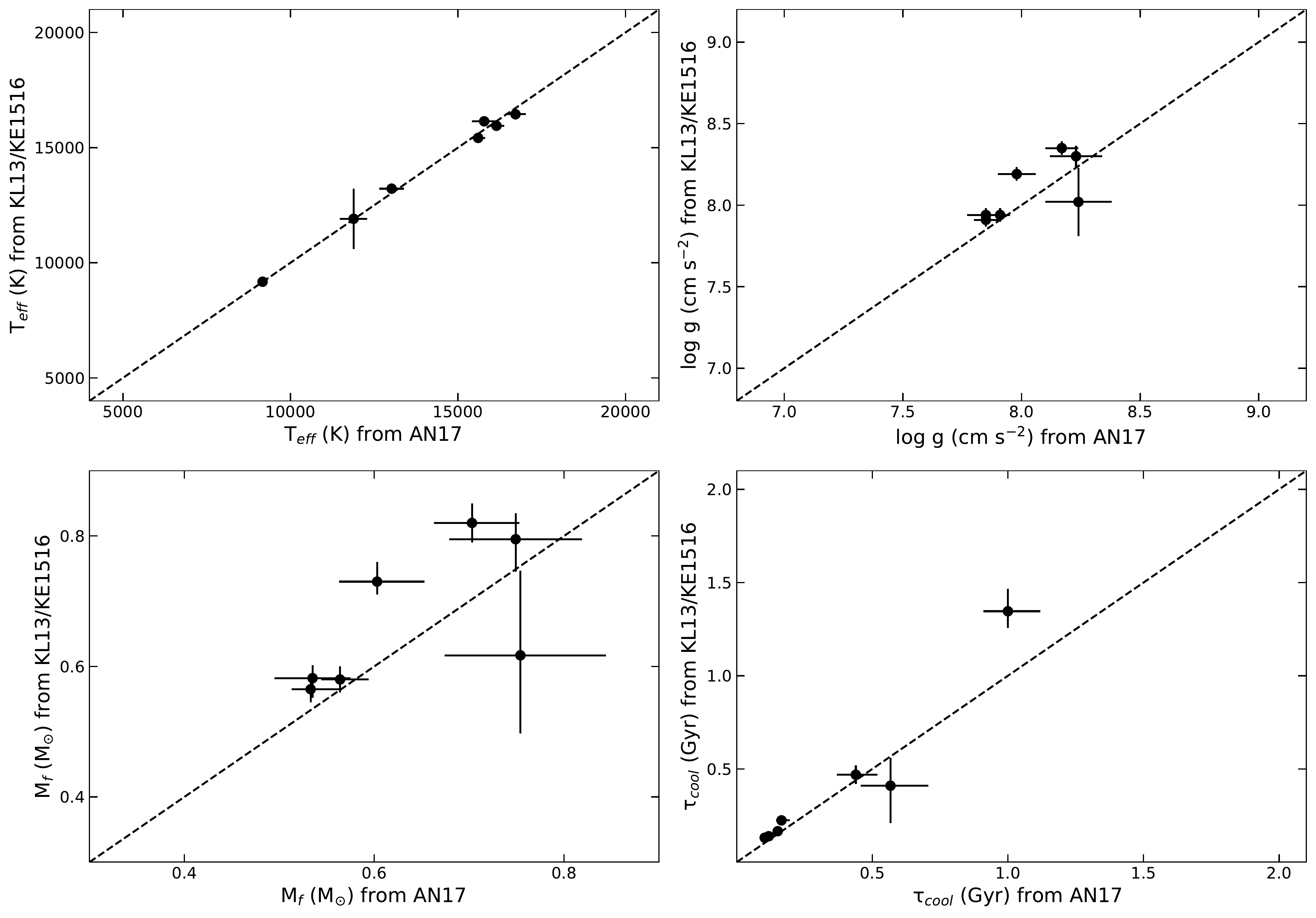}
    \caption{Spectroscopic Effective temperatures (top left panel) and surface gravities (top right panel) comparison. References used were KL13 \citep{kleinman2013}, KE1516 \citep{kepler2015,kepler2016} and AN17 \citep{anguiano2017}. Masses (bottom left panel) and cooling times (bottom right panel) comparison between the results using  AN17 and KL13/KE1516 stellar parameters with \cite{bedard2020} cooling tracks.}
    \vskip 5mm
   \label{fig:3}
\end{figure*}

We considered those pairs where the WB type was ``MS-WD'' and the primaries lay inside our TO/SG selection region, finding 54 possible systems. Of these, only 8 have stellar parameters for both components, and the secondaries are DA WDs. Effective temperatures and metallicities for the TO/SG stars were taken from catalogs such as LAMOST \citep{lamost2019}, RAVE \citep{rave2017}, and PASTEL \citep{soubiran2016}. For the WDs, we used the photometric stellar parameters reported in \cite{gf2019} (hereafter GF19). No spectra were available in SDSS for these systems.

Our final sample is composed of 15 pairs. The positions of these WBs in the CMD are shown in Figure~\ref{fig:1}, where the selected sample is represented by the linked red dots, the linked black dots are the 73 possible SG-WD pairs (Our search + SG-WD ER18), and the grey background are stars from Gaia DR2 \citep{gaiadr2} at distances $<$100 pc.

The astrometric parameters for the selected WBs are shown in Figure~\ref{fig:2}, using ER18 catalog as background for the sake of comparison. We notice that all of our pairs lay in the dense region of ER18 binaries with $\Delta V_{\perp}< 2$ km s$^{-1}$ and $\Delta \varpi < 0.5$ mas, the region of parameter space where true, gravitationally bound wide binaries are expected to live.

The complete set of proper motions, parallaxes, magnitudes, angular separations, transverse velocity differences, and projected separations for this sample are shown in Table~\ref{tab:1}.

\section{White Dwarfs Analysis}
\label{sec:3}

\subsection{Effective Temperatures and Surface Gravities}
\label{sec:3.1}

\subsubsection{Spectroscopic Parameters}
\label{sec:3.1.1}

We recompiled effective temperatures (T$_{\text{eff}}$) and surface gravities (log g) in the literature for the WDs in this work. Of our 15 WDs, only 7 of them have SDSS spectra (WDs selected in our search). The determination of these atmospheric parameters is usually accomplished by comparing predicted fluxes from model atmosphere calculations with spectroscopic data.

\cite{anguiano2017} (hereafter AN17) reported atmospheric parameters for these WDs using SDSS spectra, performing a fit of the observed Balmer lines to WDs models of \cite{koester2010}, following the procedure described in \cite{bergeron1992}. In order to account for the higher dimensional dependence of convection, which is important for cool WDs, AN17 applied the three-dimensional corrections of \cite{tremblay2013}. 

\cite{kleinman2013} and \cite{kepler2015,kepler2016} (hereafter KL13 and KE1516, respectively) also reported these parameters by using the WD's spectra and photometry in SDSS. The method used in these works is described in \cite{kleinman2004} and \cite{eisenstein2006}. In brief, they compare the spectrum and photometry of each WD candidate to a model atmosphere grid performing a $\chi^2$ minimization. By doing this, they obtained T$_{\text{eff}}$ and log g, as well as a classification of helium or hydrogen atmosphere. Tridimensional convection corrections reported in \cite{tremblay2013} were applied only in KE1516; KL13 surface gravities may differ from the other references because of this, specially for WDs with effective temperatures lower than 12000 K. 

There are some cases were the uncertainties in effective temperatures and surface gravities are very small given the high quality of the data. In order to not underestimate these errors, we will consider the scatter in spectroscopic WD parameters of $\sim$1.4 \% in T$_{\text{eff}}$ and 0.042 in log g \citep{liebert2005,bedard2017}.

The comparison of T$_{\text{eff}}$ and log g for both references is shown in the top panels of Figure~\ref{fig:3}. For most of the WDs, the difference in T$_{\text{eff}}$ is smaller than 350 K around the 1:1 line, and the log g difference is smaller than 0.2 dex. Surface gravity comparison does not look as sharp as the T$_{\text{eff}}$ because of the difference in the convection treatment by KL13 as we mention before. The larger uncertainties in some of the atmospheric parameters are because of the low signal-to-noise (S/N) of SDSS spectra. The S/N per pixel of the these WDs varies between 15 and 47 at 4700 \text{\AA }. Despite this, the measurements are consistent within the errors.

\subsubsection{Photometric Parameters}
\label{sec:3.1.2}

For the 8 remaining WDs in our sample there are not SDSS spectra available. Therefore, we used photometric atmospheric parameters available in GF19. This catalog is composed of 486641 stars with reported T$_{\text{eff}}$, log g, and masses. In brief, to calculate these quantities, the available WD magnitudes are converted into average fluxes in the different \textit{Gaia} bandpasses. After this, with \textit{Gaia} parallaxes and adequate cooling sequences, the observed fluxes are compared to model fluxes via $\chi^2$ minimization, obtaining the desire stellar parameters and their uncertainties. The latter are obtained from the covariance matrix associated with the fitting procedure \citep[e.g.][]{kilic2019}.

The T$_{\text{eff}}$ ranges from 4650 to 16470 K and log g from 7.82 to 8.40. The uncertainties in log g are above 0.10 dex in most cases, mainly because of the broad \textit{Gaia} bandpasses (G, G$_{BP}$, G$_{RP}$) and uncertainties in parallax. This will create a large scatter at the moment of determining masses and cooling times for the WDs.

\subsection{Masses and Cooling Times}
\label{sec:3.2}

Masses and cooling times for AN17, KL13, KE1516, and GF19 were found interpolating T$_{\text{eff}}$ and log g into DA theoretical cooling tracks from \cite{bedard2020}\footnote{Cooling tracks available online at \url{http://www.montrealwhitedwarfdatabase.org} \citep{dufour2017}.}. These cooling tracks assumed that a typical WD can be represented by DA models with standard ``thick'' hydrogen layer containing 10$^{-4}$ of the total mass of the star (M$_{H}$/M$_{\odot}$= 10$^{-4}$).

The comparison between the values calculated using both sets of spectroscopic stellar parameters can be seen in the lower panels of Figure~\ref{fig:3}. For almost all the WDs, the differences in M$_{\text{f}}$ are smaller than 0.14 M$_{\odot}$ with uncertainties better than 0.10 M$_{\odot}$. This scatter in M$_{\text{f}}$ is expected for cool WDs, as is the case of WD3 and WD7 because of the high log g problem given the unidimensional treatment of convection. This will also lead to differences under 0.13 Gyr in t$_{\text{cool}}$. 

The T$_{\text{eff}}$, log g, WD masses (M$_{\text{f}}$), cooling times (t$_{\text{cool}}$), and their uncertainties are listed in Table~\ref{tab:A.2} for all the references in this work.

For our purposes, we need the best possible precision to constrain the IFMR. Both spectroscopic references and the photometric reference shown in this Section are equally valid, therefore we will consider them all and determine later what is the impact, if any, on the constraints for the IFMR.

\smallskip
\section{Turnoff/Subgiant Companion Analysis}
\label{sec:4}

\subsection{Observations and Data Reduction}
\label{sec:4.1}

We managed to obtain spectra of the 7 primaries in Table~\ref{tab:1} during three different observing runs 2017 July 25, 2018 October 01, and 2019 January 03\footnote{The data was obtained under the CNTAC program ID CN2017B-85, PI M. Barrientos.}. All the stars were observed with the high-resolution spectrograph  Magellan Inamori Kyocera Echelle \citep[MIKE;][]{bernstein2003} on the 6.5 m Clay Telescope at Las Campanas Observatory. We used a narrow slit ($0.35''$), which delivers data with spectral resolution R = $\lambda/\Delta\lambda$ $\simeq$ 65000 (at $\lambda$ $\simeq$ 6000 \AA) and the standard setup that allows complete wavelength coverage in the 3400-9100 \AA \text{ } spectral window. The S/N per pixel of the observed stars varies between 100 and 300 at 6000 \text{\AA }. The spectra were reduced using the CarPy pipeline\footnote{\url{https://code.obs.carnegiescience.edu/mike}}, which trims the spectra and corrects for overscan, apply both milky and quartz flats, removes scattered light, and subtracts sky background. Then, the stellar flux is extracted order-by-order and a wavelength solution is applied using a wavelength map and the Th-Ar lamps taken during the observing night. Using stardard tasks from IRAF\footnote{IRAF is the Image Reduction and Analysis Facility, a general-purpose software system for the reduction and analysis of astronomical data. IRAF is written and supported by National Optical Astronomy Observatories (NOAO) in Tucson, Arizona.}, we proceed to continuum normalize the observed spectra order-by-order and merge it. Thereafter, we used the Python package PyAstronomy within \texttt{crosscorrRV} task to correct the spectra for radial velocity using a cross-correlation function and a radial velocity standard. The accuracy for these measurements is $\sim0.5$ km s$^{-1}$, which is sufficient for our scientific purposes.

\subsection{Determination of Atmospheric Parameters}
\label{sec:4.2}

In order to determine the age of the TO/SG in our WBs, first, we need to measure atmospheric parameters such us T$_{\text{eff}}$, [Fe/H]\footnote{We use the standard notation: [Fe/H] = A$_{\text{Fe}}^{*}$ - A$_{\text{Fe}}^{\odot}$, where A$_{\text{Fe}}$ = log(N$_{\text{Fe}}$/N$_{\text{H}}$) + 12 and N$_{x}$ is the number density of X atoms in the stellar photosphere.}, and log g.

We obtained these fundamental atmospheric parameters for SG1 to SG7 by using the standard iron line spectroscopic approach, which forces excitation and ionization balance of Fe I and Fe II lines. We first selected an average of 100 iron lines (86 Fe I and 14 Fe II) employed in \cite{ramirez2014} with atomic data available and equivalent width (EW) measurements of a reference Sun (asteroid) spectrum. Thereafter, for every observed primary, we measured ``manually'' the EW of the iron lines by fitting Gaussian profiles using the IRAF's task \texttt{splot}. Then, a first guess of the parameters is made for each observed star. In this case, we took the Sun as a reference because of the similar spectral type and chemical properties with our sample of stars. Considering both requirements, we used the spectrum synthesis code MOOG \citep{sneden1973}\footnote{\url{ http://www.as.utexas.edu/~chris/moog.html}} and the q$^2$ python package\footnote{Package developed by Ivan Ramirez. The source code is available online at \url{https://github.com/astroChasqui/q2}}, which removes the complexities in the manipulation of input and outputs of MOOG to calculate iron abundances of Fe I and Fe II on a line-by-line basis. For this process, we adopted atmospheric models from MARCS grid of standard chemical compositions \citep{gustafson2008}\footnote{\url{https://marcs.astro.uu.se/}}. 

\begin{figure}[t!]
	\includegraphics[width=0.45\textwidth]{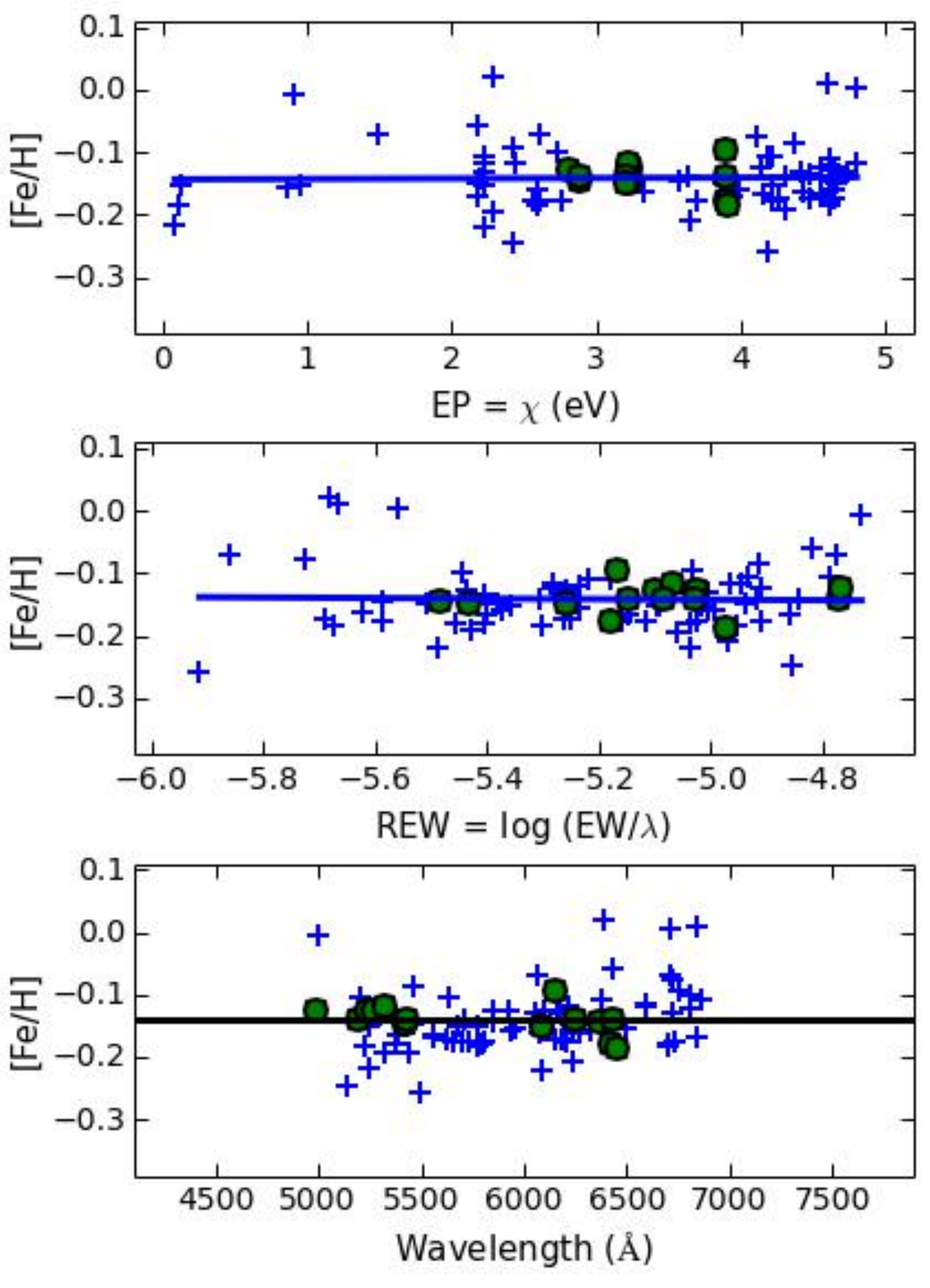}\\
    \caption{An example of MOOG's iterative procedure output. Line-to-line relative iron abundance as a function of excitation potential (top panel), reduced equivalent width (middle panel), and wavelength (bottom panel). Crosses (circles) are Fe I (Fe II) lines. The solid lines in the top and middle panels are linear fits to the Fe I data. In the bottom panel, the solid line is a constant which corresponds to the average iron abundance}
   \label{fig:4}
\end{figure}

MOOG's iterative process refine the atmospheric parameters for each star by removing the correlation of relative iron abundance (with respect to the Sun) with excitation potential (EP), thus forcing the excitation balance (related to the T$_{\text{eff}}$). Besides, the correlation between the intensity of the lines, i.e, the reduced equivalent width (REW= log [EW/$\lambda$]) and the relative iron abundance is controlled by the microturbulence velocity (v$_t$). Both of these correlations can be seen in Figure~\ref{fig:4} that is part of the end product of MOOG's process. At the same time, this tool eliminates the difference between mean abundances of Fe I and Fe II obtained separately, thus reaching ionization balance (related to log g). The [Fe/H] adopted was calculated as the average in each iteration. The errors for these parameters were obtained propagating the uncertainties in the iron abundance versus EP and REW slopes, added in quadrature with the errors from the line-to-line scatter between mean Fe I and Fe II abundances. For [Fe/H], the error was also calculated adding in quadrature the standard deviation of the line-to-line dispersion \citep[see][]{ramirez2014,navarrete2015}.

For SG5, we could not measure EWs because of readout problems in the spectrum of the star. Thus, we will not use this pair (N\textdegree 5 in Table~\ref{tab:1}) to constrain the IFMR, reducing our sample to 14 pairs. 

For SG8 to SG15, we obtained the stellar parameters from different catalogs in the literature such as \cite[LAMOST][]{lamost2019}, \cite[RAVE][]{rave2017}, and \cite[PASTEL][]{soubiran2016}. If more than one reference per star was available, we used the results from the best data in terms of S/N.

All the derived and literature stellar parameters along with their uncertainties are listed in Table~\ref{tab:A.3}. 

\begin{figure*}
\centering
	\includegraphics[width=0.95\textwidth]{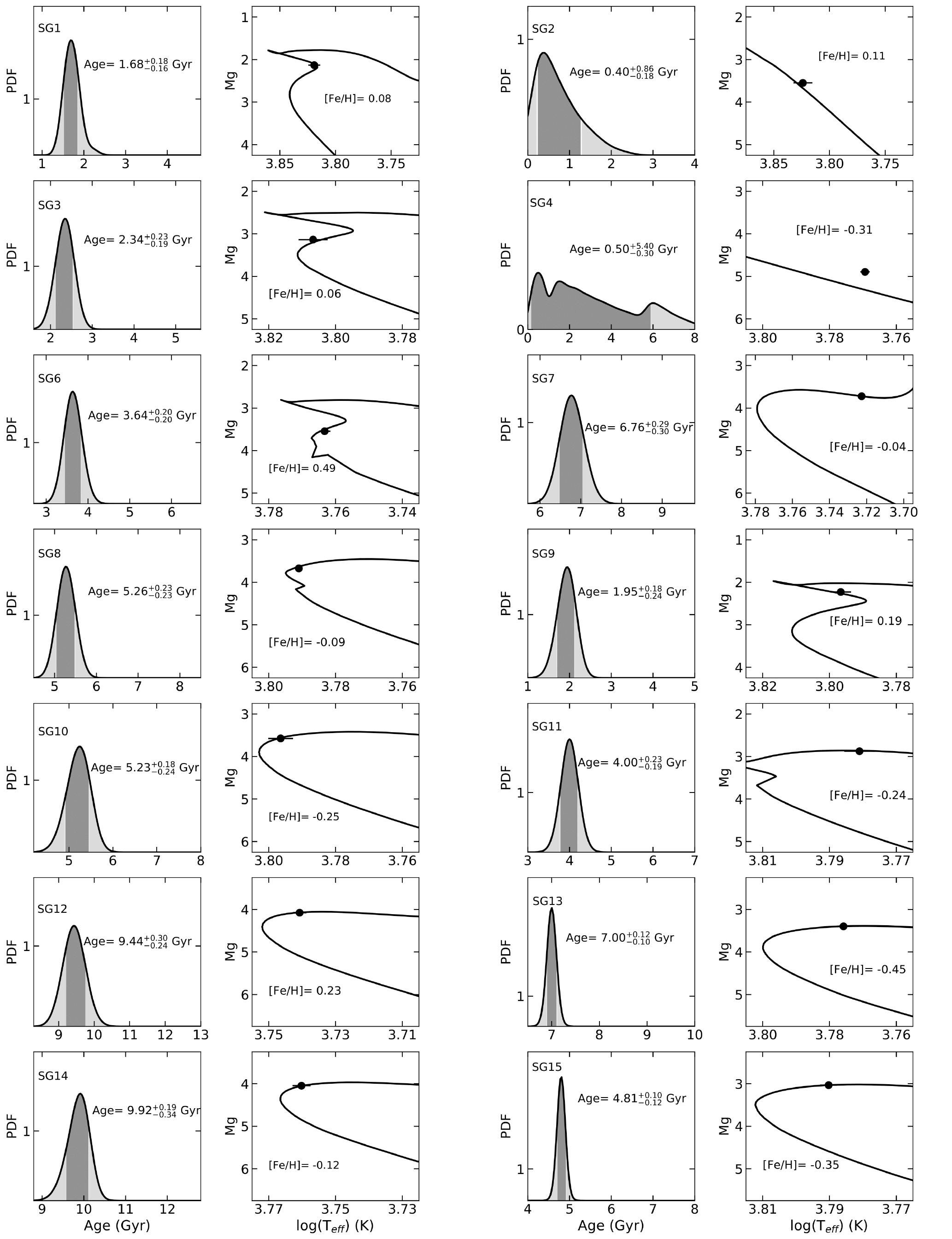}
    \caption{Left panel: Age PDF for each primary in our sample using \texttt{MIST} isochrones. The peak of the distribution is the most probable age of a star, and the transition between dark and light grey show the upper (84\%) and lower (16\%) 1$\sigma$ error. Right panel: Isochrone of the most probable age fitted to the primary star using its metallicity.}
   \label{fig:5}
\end{figure*}

\subsection{Age Determination}
\label{sec:4.3}

The ages for our full sample of primaries were computed by using theoretical isochrones. In this technique, the star under study is placed on a CMD and its location compared to theoretical predictions of stellar evolution. Isochrone points close to the observed stellar parameters are then used to derive the age of the star \citep[e.g.,][]{lachaume1999,nordstrom2004,jl2005,cr12,ag2018}. While the location of any star on the CMD is determined just by its absolute luminosity (e.g., Mv) and T$_{\text{eff}}$ (or color), the [Fe/H] must also be given in order to infer the age of the star avoiding degeneracies \citep[e.g.,][]{casagrande2011}. 

Typically, the fundamental atmospheric parameters T$_{\text{eff}}$, log g, and [Fe/H] are measured from the star’s spectrum, as we did in Section~\ref{sec:4.2}, and they are enough to determine the age of the star with this method. In our case, we have exquisite parallaxes measured by \textit{Gaia}, thus we decided to use the absolute magnitude of the star which is determined by its apparent magnitude and parallax instead of surface gravity.

\begin{figure}[t!]
\begin{center}
	\includegraphics[width=0.47\textwidth]{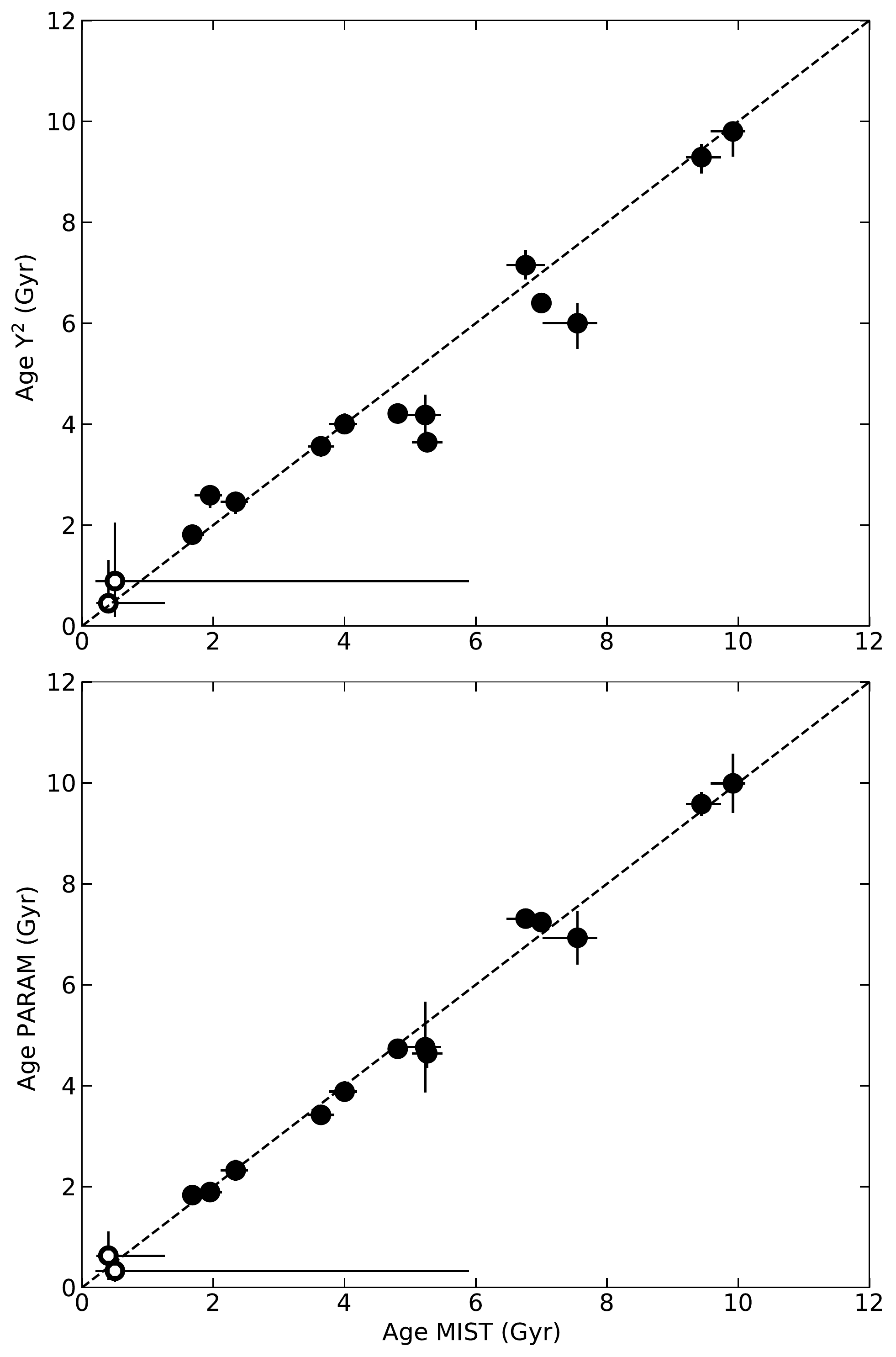}\\
    \caption{Comparison of ages using \texttt{MIST} isochrones \citep{dotter2016,choi2016}, \texttt{Y}$^2$ isochrones \citep{yi2001,demarque2004}, and PARAM \citep{dasilva2006}. Black dots are TO/SG primaries and the open circles are MS primaries.}
   \label{fig:6}
\end{center}
\end{figure}

In practice to calculate the age of the stars, we create a probability distribution function (hereafter PDF) of the stellar age by comparing the stellar parameters and uncertainties of T$_{\text{eff}}$, Mv, and [Fe/H] to isochrones passing near this set of parameters, following the procedures described in \cite{cr12}. The PDF obtained was smoothed and normalize using a Gaussian kernel density estimate. Finally, the peak of the PDF was selected as the most probable age of a star. The lower and upper 1$\sigma$ errors for the age were determined as the 16\% and 84\% of the cumulative probability, respectively.

Using this procedure, we computed the ages of all primaries in our sample using MESA Isochrones \& Stellar Tracks isochrones \citep[\texttt{MIST};][]{dotter2016,choi2016}. We also used Yale-Yonsei isochrones \citep[\texttt{Y}$^2$;][]{yi2001,demarque2004} for comparison. Both isochrone sets have an age range of 0.1 to 14 Gyr in steps of 0.1 Gyr. While creating the age PDF (as was mention above), we tested bin sizes from 0.1 to 1 Gyr, finding that 0.2 Gyr/bin was an ideal choice. The ages and uncertainties for all the primaries in our sample are listed in Table~\ref{tab:A.3}. 

Figure~\ref{fig:5} show the resulting age PDFs and 1$\sigma$ uncertainties around the most likely age (left panel) and the isochrones fitted to stars (right panel) using \texttt{MIST} isochrones. From the right panels of Figure~\ref{fig:5}, we can notice that some of our stars are not TO/SG stars. This is the case of SG2 and SG4 (MS stars). We separated this group of stars from the rest of the sample, and they will not be used to constrain the IFMR. The top panel of Figure~\ref{fig:6} shows the comparison between the ages determined using both isochrone sets, including the non TO/SG primaries. Results from both isochrone sets show consistency, with a dispersion below $\sim$0.30 Gyr for most of our sample.

In order to account for systematic biases in the isochrone technique \citep[see][]{jl2005,dasilva2006}, we also compute Bayesian ages using the web interface \texttt{PARAM}\footnote{\url{http://stev.oapd.inaf.it/cgi-bin/param_1.3}} described in \cite{dasilva2006}. This public tool uses PARSEC isochrones \citep{bressan2012} and differs from the standard isochronal approach because requires the selection of priors for the analysis. We choose an initial mass function from \cite{chabrier2001} and a constant star formation rate from 0.1 to 14 Gyr. The resulting ages and their uncertainties are also listed in Table~\ref{tab:A.3}. The age comparison between \texttt{MIST} and \texttt{PARAM} is shown in the lower panel of Figure~\ref{fig:6}. Despite the fact that these Bayesian ages are obtained in a manner completely independent from our own isochrone analysis, we can see that both ages and uncertainties are remarkably similar, around the 1:1 line even for the oldest star in our sample. This reflects the fact that the subgiant branch is a privileged evolutionary stage where isochrone ages are reliable, have high time-resolution and relative precision \citep[e.g.,][]{gr2021}.

\subsubsection{Age Comparison: Isochrones vs Asteroseismology}
\label{sec:4.3.1}

In order to check our results and the isochrone fitting technique, we selected a sample of SG stars with asteroseismic ages in \cite{chaplin2014} and \cite{tandali2020} using spectroscopic stellar parameters (T$_{\text{eff}}$ and [Fe/H]) from \cite{bruntt2012}, \cite{bl2015}, and \cite{Mathur2017}. 

\begin{figure}[t!]
\begin{center}
	\includegraphics[width=0.47\textwidth]{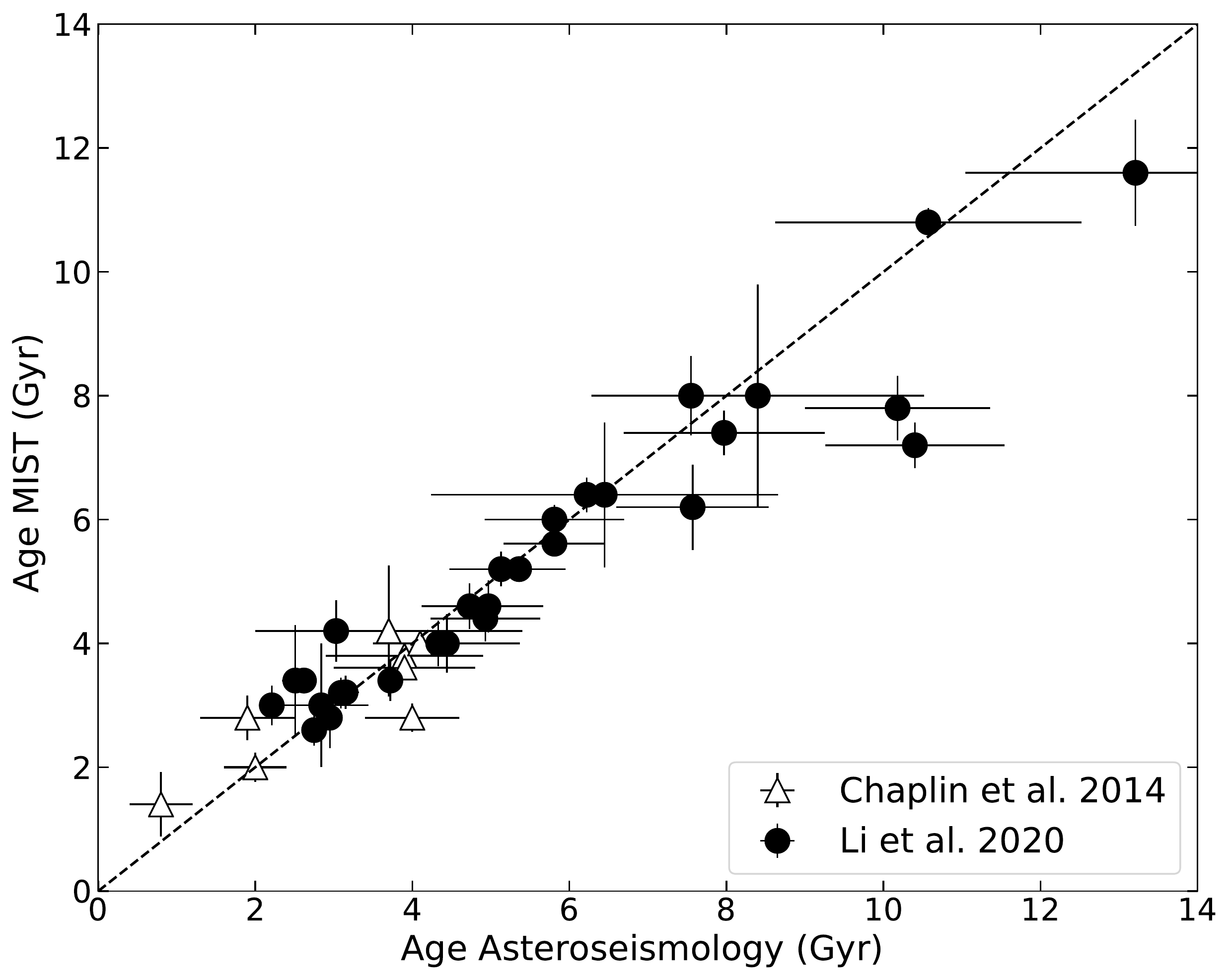}\\
    \caption{Comparison of ages as determined with our procedure with \texttt{MIST} isochrones and seismic ages by \cite{chaplin2014} and \cite{tandali2020} for a sample of TO/SG stars.}
   \label{fig:7}
\end{center}
\end{figure}

We calculated ages for this sample of stars via \texttt{MIST} isochrones. As we can see in Figure~\ref{fig:7}, the isochrone technique yielded similar ages as the asteroseismology method, especially between 2 and 8 Gyr; which is the age range where most of our sample lies. Over this age limit, the asteroseismic ages tend to be slightly older than isochrone ages. Therefore, the good agreement with seismic ages below 8 Gyr provides additional independent confidence to our own isochrone ages.

\section{The Initial-to-Final Mass Relationship}
\label{sec:5}

\subsection{Initial Mass Determination}
\label{sec:5.1}

Once we have the age of the system (i.e, the TO/SG age), the final step to constrain the IFMR is to calculate the progenitor's lifetime for each WD in our pairs comparing the system ages and cooling times as follows 
\begin{equation}
\label{eq:10}
\begin{split}
\text{Age}_{\text{TO/SG}} = \text{Total Age}_{\text{WD}} = \tau_{\text{cool}} + \tau_{\text{prog}} \textrm{ , } \\
\tau_{\text{prog}} = \text{Age}_{\text{TO/SG}} - \tau_{\text{cool}} \textrm{. }
\end{split}
\end{equation}

\noindent With these values and the metallicities of their TO/SG companions we can infer the initial masses using stellar evolution models. To be consistent with the age determination, we used \texttt{MIST} evolutionary tracks to create progenitor lifetime functions (hereafter PLF) in a range of initial masses from 0.8 to 8.0 M$_{\odot}$ with different metallicities from +0.50 to -0.50 dex. We define as progenitor lifetime the time spent from the zero-age main sequence to the first thermal pulse on the AGB. This choice avoids having to deal with the complexities associated to stellar evolution during thermal pulses, and is justified in that the duration of this evolutionary phase is quite rapid ($\sim 10^6$ yr) in comparison to the $\gtrsim$ Gyr total ages of the stars we are concerned with in the present work \citep[see][]{andrews2015}.

\begin{figure}[t!]
	\includegraphics[width=0.47\textwidth]{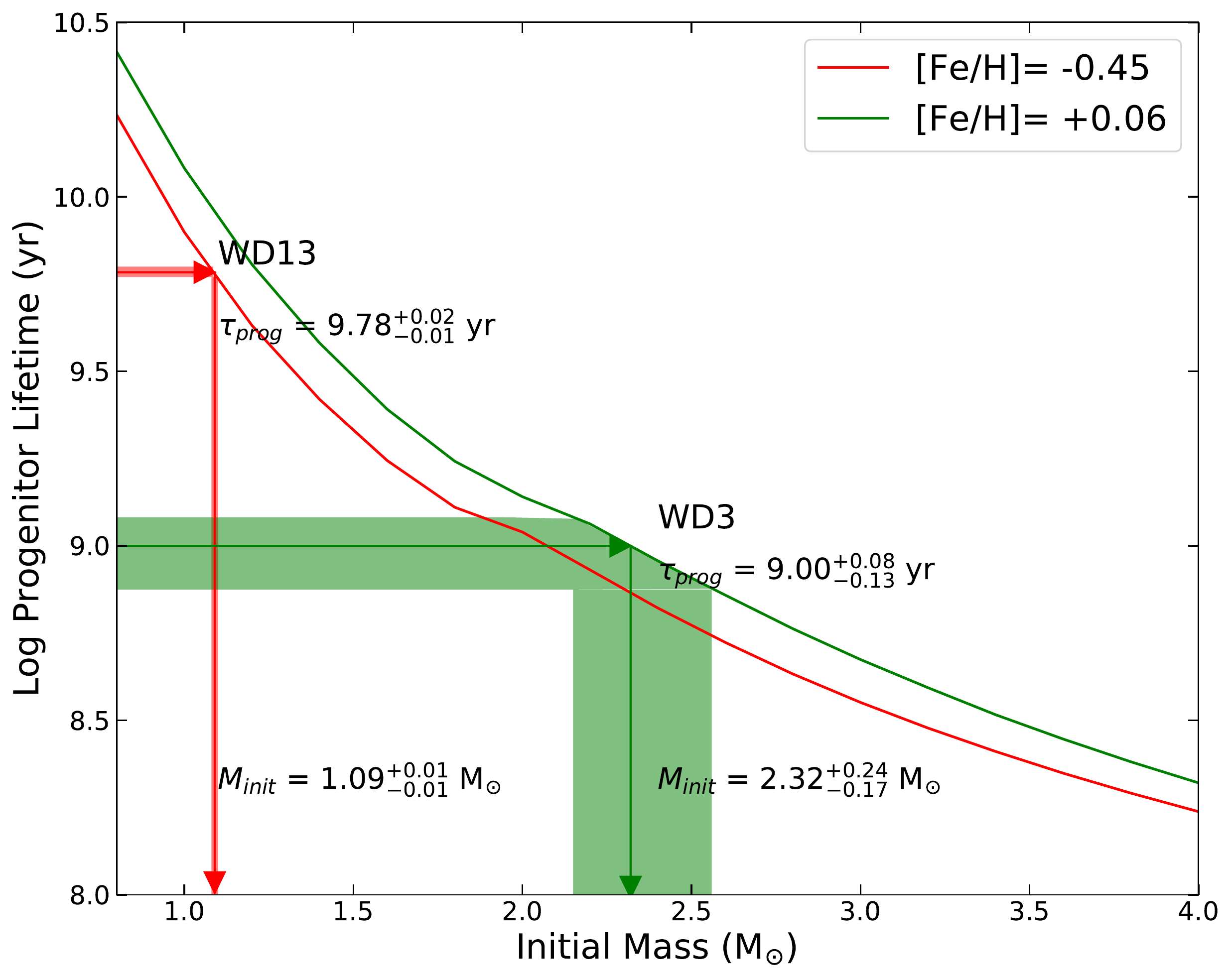}
    \caption{An example of how the progenitor lifetimes are converted into initial masses using evolutionary tracks. The red and green arrows indicate the progenitor lifetimes mapped into initial masses as obtained using MIST stellar models with [Fe/H]= -0.45 and [Fe/H]= 0.06 PLFs, respectively. The red and green area indicate the errors for each measurement, for the two WDs chosen to illustrate the procedure.}
   \label{fig:8}
   \smallskip
\end{figure}

As an illustration, Figure~\ref{fig:8} shows the mapping of 2 WD's progenitor lifetimes into their progenitor masses for a given metallicity. The progenitor lifetime ($\tau_{\text{prog}}$) of all the WDs studied here and their initial masses (M$_{\text{i}}$) are listed in Table~\ref{tab:A.2}.

\subsection{Constraining the Low-Mass End of the IFMR}
\label{sec:5.2}

\begin{figure*}[t!]
\centering
	\includegraphics[width=\textwidth]{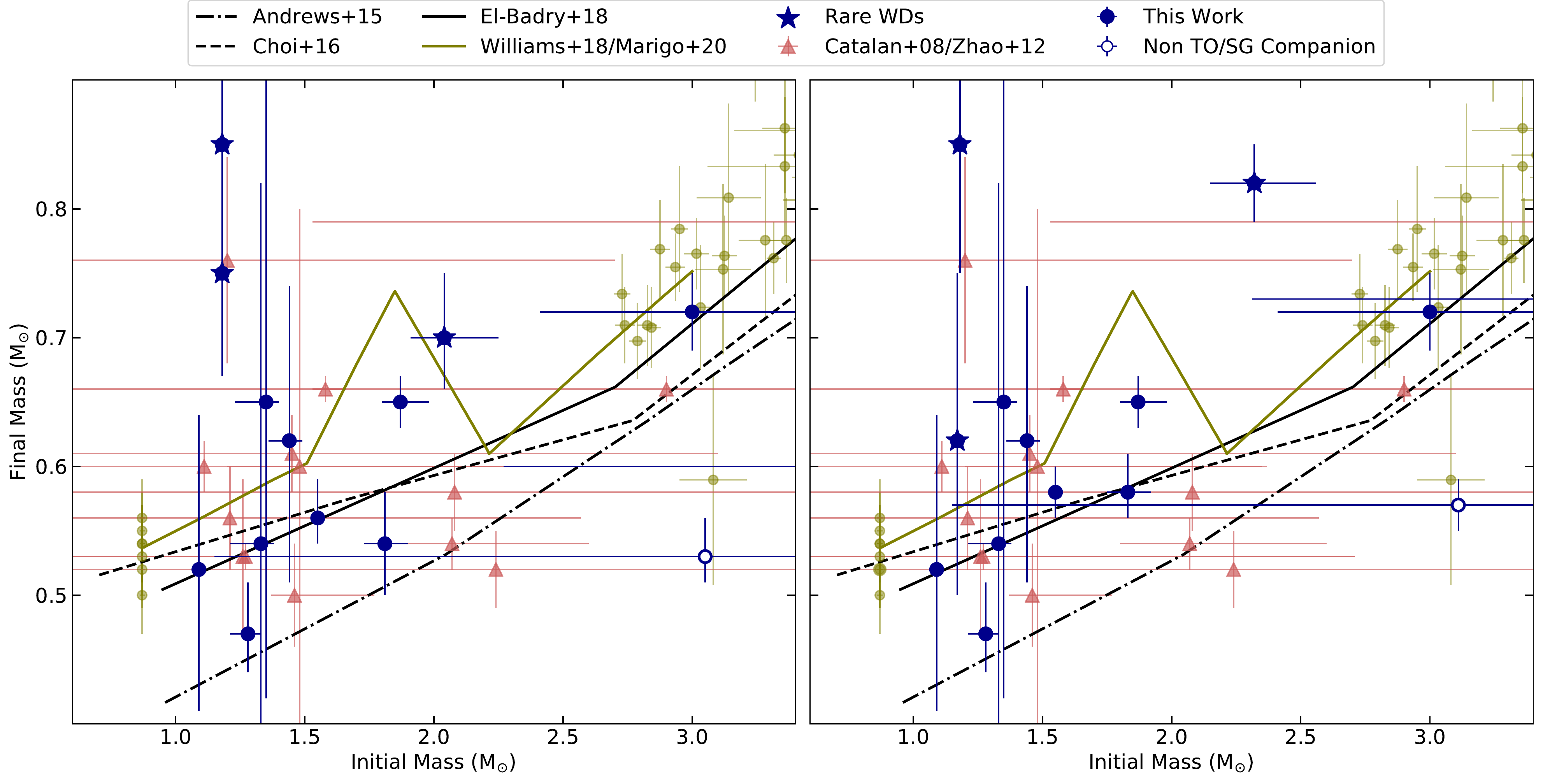}
    \caption{Constraints for the IFMR obtained in this work, using the stellar parameters from AN17 (left panel), KL13, and KE1516 (right panel), and photometric parameters from GF19 (both panels). Red triangles are WB constraints from \cite{catalan2008a} and \cite{zhao2012}, green dots and solid line are OCs constraints from \cite{williams2018} and \cite{marigo2020}, blue circles are the WB constraints obtained in this work, from TO/SG primaries (filled circles), and from non-evolved primaries (open circles). Blue stars are our rare WDs with high final masses. The dashed line in black is the theoretical IFMR by \cite{choi2016}, the black solid line is the IFMR by \cite{elbadryb2018}, and the \cite{andrews2015} relation is the black dot-dashed line.}
   \label{fig:9}
\end{figure*}

Figure~\ref{fig:9} displays the resulting semi-empirical constraints for the IFMR using 11 WDs in our sample (filled and open circles). We choose to show our results for two sets of spectroscopic final (WD) masses in two corresponding panels, one set obtained using WD stellar parameters from AN17 (left panel), and another reported by KL13, KE1516 (right panel). Both panels also show final masses obtained using photometric parameters from GF19 (see Section~\ref{sec:3.1}). The difference between filled and open circles comes from the age determination. Open circles are constraints obtained from non TO/SG companions. For the sake of comparison we have also included the WDs in WBs from \cite{catalan2008a} (hereafter CA08) and \cite{zhao2012} (hereafter ZH12) as red triangles, the WDs in stellar clusters from \cite{williams2018} (hereafter WI18) and \cite{marigo2020} (hereafter MA20) as the green solid line and dots, semi-empirical IFMRs by \cite{andrews2015} (hereafter AN15) and \cite{elbadryb2018} (hereafter EB18b) as the black dot-dashed and solid lines, respectively, and a theoretical IFMR from \cite{choi2016} (hereafter CH16) as the black dashed line.

All the constraints found in this work have M$_{\text{i}}$ $<$ 3.0 M$_{\odot}$, allowing us to better constrain the IFMR in the low-mass end, which is a difficult limit to reach using WDs in stellar clusters. This region was already accessible with the WBs studied by CA08 and ZH12. Comparing our results with CA08 and ZH12, their uncertainties in the progenitor mass are significantly larger than ours. This is because of the poor accuracy obtaining ages for MS stars using isochrones. For MS stars with ages between 0.5 and 6.5 Gyr, the uncertainties are between 30\% and 50\% in most of their sample. Consequently, the errors in M$_{\text{i}}$ measurements are between 14\% and 90\% for stars with 1.1 $<$ M$_{\text{i}}$/M$_{\odot}$  $<$ 4.1. In contrast, the TO/SG stars in our sample with ages between 1.7 and 9.9 Gyr have errors below 10\%, yielding M$_{\text{i}}$ uncertainties between 2\% and 10\% for stars with 1.1 $<$ M$_{\text{i}}$/M$_{\odot}$ $<$ 2.3, improving considerably the previous M$_{\text{i}}$ determination using WBs to study the IFMR.

One of the main contributions of the present work, therefore, is that the larger error bars in the low-mass region of Figure~\ref{fig:9} are not anymore from the initial mass determinations (M$_{\text{i}}$), but they are now from the (final) masses of the WDs, mostly due, to the low S/N of the SDSS spectra they depend on; the large uncertainties in (some WDs) parallaxes and the broad \textit{Gaia} bandpasses in the case of the photometric parameters. Under this light, we can say that the precision of empirical constraints on the IFMR in the low-mass regime are now limited by the final mass determination (M$_{\text{f}}$). With the use of WBs containing a WD and a TO/SG companion, we are not limited anymore by the precision in M$_{\text{i}}$, but now the pressure is on the measurement of the WD masses today.

The constraints from WDs in stellar clusters presented by WI18 and MA20 covers a similar region than our constraints in the IFMR, with around 18 WDs members from M67, NGC 6819, R-147, NGC 752, and NGC 7789 between 1.3 and 2.1 M$_{\odot}$. In Figure~\ref{fig:9}, we represent these WDs with a green solid line. We notice that our constraints are spread over and under the traced relation with OCs WDs between 1.0 $<$ M$_{\text{i}}$/M$_{\odot}$ $<$ 1.5. However, we can see an apparent increase in  M$_{\text{f}}$ for 1.5 $<$ M$_{\text{i}}$/M$_{\odot}$ $<$ 2.0. Better determinations of M$_{\text{f}}$ are necessary to confirm the ``kink'' behaviour in our sample. In general, stellar clusters produce more massive WDs for a given M$_{\text{i}}$, in comparison with semi-empirical IFMRs by AN15, EB18b, and the theoretical IFMR by CH16. The differences with theoretical IFMRs was analyzed by \cite{cummings2019}, suggesting that the discrepancy in M$_{\text{f}}$ at low masses (M$_{\text{i}}$ $<$ 3 M$_{\odot}$), results from the handling of the models in the AGB phase.

Inspection of Figure~\ref{fig:9} reveals that the new WB constraints, if taken at face value, suggest a large dispersion in the final masses in the low-mass regime of M$_{\text{i}} \sim$ 1.0-1.5 M$_{\odot}$. The dispersion does not seem to be accounted for by the uncertainties in M$_{\text{f}}$, regardless of which set of final masses we use, and is in large part driven by three WDs (WD3, WD7, and WD12) that would seem overmassive given their progenitor masses.  In what follows, we will see that two of these have been shown to be close double degenerates, which means that, along with their distance SG companions, they comprise not only wide binaries but hierarchical triples.  Excluding these (double) WDs from the IFMR analysis, however, reduces the apparent dispersion but does not get rid of it. If confirmed real, this dispersion would have important implications for the IFMR and its application to a variety of problems in astrophysics. These WDs are thus very special targets for this field and should be followed up with improved data and analysis.

Given their importance, next we consider the possibility that something went wrong in our procedure for these three WDs. We can think of different scenarios that would invalidate the use of these three objects for studying the IFMR:
\begin{enumerate}
    \item The WD masses are overestimated: however, the dispersion created by these potential outliers does not seem to be explained by the fact that KL13 did not apply 3D corrections in the convection treatment. Indeed, as the left panel of Figure~\ref{fig:9} shows, use of the M$_{\text{f}}$ obtained by AN17 accounting for these 3D corrections result in the same dispersion in this low-mass regime. Another possibility for the characterization of the masses of these WDs is through the measurement of their gravitational redshifts, which we attempt and discuss in Section~\ref{sec:5.3.1} below.
    
    \item These WDs are double degenerate systems. This type of binaries can be identify through the analysis of the luminosity of WDs, and also studying the inconsistencies between spectroscopic and photometric solutions \citep[e.g.,][]{kilic2021a}. We will discuss this possibility in Section~\ref{sec:5.3.2}.
    
    \item These WDs are merger remnants or remnants of blue stragglers. There is evidence that an important part of the massive WDs in our neighbourhood likely formed through mergers in binary systems \citep[e.g.,][]{kilic2018,kilic2021b}. At the same time, close binary interaction can produce blue stragglers with larger masses, hence more massive WDs \citep[e.g.,][]{williams2018,Casagrande2020}. Considering the M$_{\text{f}}$ of our outliers around M$_{\text{i}}\sim$ 1.18 M$_{\odot}$, this scenario is possible.
    
    \item  The primaries to these three WDs are in reality significantly younger than the age determined in this work and reported in Table~\ref{tab:A.3}. If this were the case, then the masses of these WDs would not necessarily be anomalous in the context of the IFMR, as the progenitor masses would be larger than those determined here.

    \item The binaries reported here with these WDs as components, are in reality not true, bona fide WBs, but a pair of unrelated stars, perhaps a chance alignment that crept in our WB search algorithm. Recalling Figure~\ref{fig:2}, the astrometric parameters of these three pairs have differences in $\Delta V_{\perp}$ and $\Delta \varpi$ below 2 km s$^{-1}$ and 0.5 mas, respectively; with projected separations below 18000 au, safely below the selection threshold.

\end{enumerate}

All this warrants a closer inspection of the outliers in our results. Because we do not have a spectroscopic reference for WD12, we will attempt the analysis only for WD3 and WD7 in Section~\ref{sec:5.3} ahead.

\begin{figure*}[t!]
\centering
	\includegraphics[width=\textwidth]{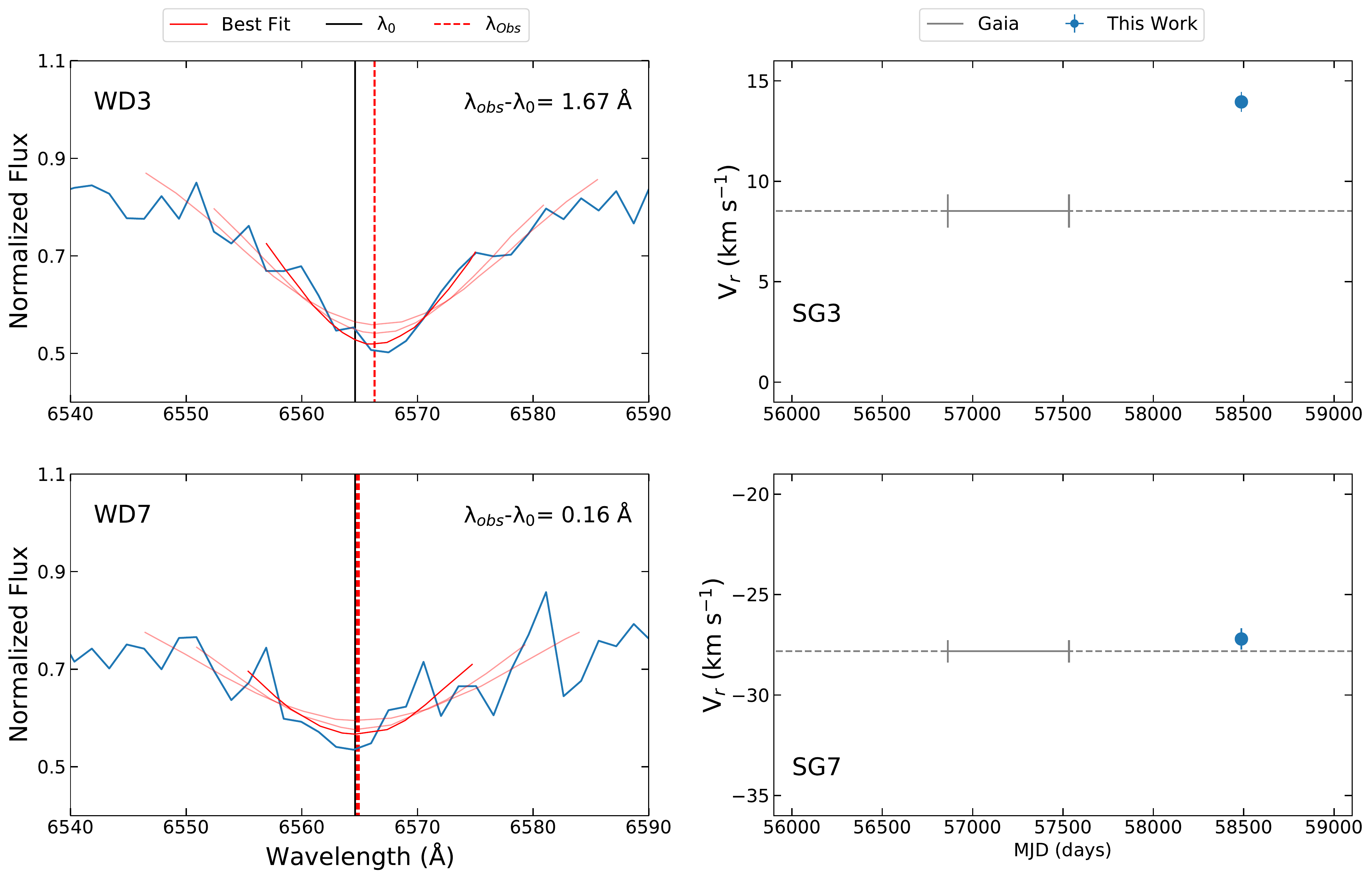}
	\caption{Left panel: Fitting procedure to obtain the V$_{\text{app}}$ of WDs. We fit the H$_{\alpha}$ line using three different windows 20 \text{\AA }, 15 \text{\AA }, and 10 \text{\AA} (Best fit, red curve). The black solid line indicates H$_{\alpha}$ rest wavelength and the red dashed line the centroid fitted. Right panel: Radial velocities (V$_r$) of the SG companions. The blue dots are the values obtained in this work and grey lines are V$_r$ by \textit{Gaia} \citep{gaiadr2}}
   \label{fig:10}
\end{figure*}

\subsection{Inspection of Rare WDs in Our Sample}
\label{sec:5.3}

Here we perform a closer inspection to two of the three massive WDs that would contribute to a dispersion in the IFMR at low masses in Figure~\ref{fig:9}. This will take us to consider the nature of the evolved primaries to these WDs, and the WDs themselves, where we find evidence that they may harbor close, unseen companions. If the latter is true, it would essentially invalidate the use of these particular systems for constraining the IFMR, since the stellar parameters for both, TO/SG and WDs, may be severely compromised by the unknown effects from a close companion.
 
\subsubsection{Measuring Gravitational Redshifts}
\label{sec:5.3.1}

As was done by \cite{silvestri2001}, WDs in common proper motion pairs can be used to determine gravitational redshift masses (M$_{\text{GR}}$). We can disentangle the redshift or blueshift associated to the recession velocity from the gravitational redshift of the WD (V$_{\text{GR}}$) by using the companion (in this case the TO/SG star), which gives a relative standard of rest. The TO/SG companion provides the intrinsic radial velocity of the system (V$_{r}$) and compared to the WD apparent radial velocity (V$_{\text{app}}$), the V$_{\text{GR}}$ can be obtained as
\begin{equation}
    V_{GR} = V_{\text{app}} - V_{r} = \frac{GM}{Rc},
    \label{eq:11}
\end{equation}
where V$_{\text{app}}$ is usually measured from the shift of the H$_{\alpha}$ line centroid relative to its rest wavelength in the WD's spectra \citep[e.g.,][]{silvestri2001,falcon2010,napi2020,chandra2020}, and V$_{r}$ typically determined cross-correlating the companion's high-resolution spectra with the spectrum of a radial velocity standard star. Also, we can see from Equation~\ref{eq:11} that this measurement is proportional to the ratio between the mass and the radius of the WD. Therefore, this method relies on the precision of the mass-radius relation for DA WDs \citep[e.g.,][]{gb2019}. At the same time, it is independent of the stellar parameters of the WDs being examined and of the atmospheric models used in the determination of cooling times and mass, and this is what would help our purposes of studying the observed scatter among our IFMR constraints. 

\begin{table*}[t!]
\centering
\caption{Gravitational Redshift Measurements for Outliers}
\label{tab:2}
\def\arraystretch{1.8}
\setlength{\tabcolsep}{10pt}
\resizebox{\textwidth}{!}{
\begin{tabular}{c c c c c c c c c}
\hline\hline
Name & Gaia source ID & SN & V$_{\text{app}}$ & V$_{\text{r,hel}}$ & V$_{\text{GR}}$ & M$_{\text{GR}}$ & M$_{\text{AN17}}$ & M$_{\text{KL13,KE1516}}$ \\
 & &  & (km s$^{-1}$) & (km s$^{-1}$) & (km s$^{-1}$) & (M$_{\odot}$) & (M$_{\odot}$) & (M$_{\odot}$)\\
\hline\hline
WD3 & 668562516332550784 & 30 & 76.43 $\pm$ 18.32 & 13.95 $\pm$ 0.50 & 62.48 $\pm$ 18.33 & 0.90 $^{+0.10}_{-0.15}$ & 0.70 $_{-0.04}^{+0.05}$ & 0.82 $_{-0.03}^{+0.03}$\\
WD7 & 3975115240311668096 & 15 & 7.33 $\pm$ 17.32 & -27.21 $\pm$ 0.50 & 34.54 $\pm$ 27.33 & 0.66 $_{-0.23}^{+0.16}$ & 0.75 $^{+0.09}_{-0.08}$ & 0.62 $^{+0.13}_{-0.12}$  \\
\hline\hline
\end{tabular}}
\smallskip
\end{table*}

However, measuring V$_{\text{app}}$ with good precision is challenging, mainly because of the pressure broadening in the wings of the Balmer absorption lines \citep[see][]{halenka2015}. To isolate this effect and accurately measure the non-broadened line core of H$_{\alpha}$ and the other Balmer lines, it is necessary to have high resolution spectra of the WDs \citep[e.g.,][]{napi2020}. SDSS spectra used in this work are limited in resolution, and do not have the power to resolve the line core of the Balmer lines \citep[see][]{chandra2020}. Consequently, our measurements based on SDSS spectra will not be as precise as we need for our purposes of constraining the IFMR, but it will provide an independent comparison with the (apparently anomalously high) spectroscopic masses determined from the WD model atmospheres.

We measured V$_{\text{app}}$ for the three WDs with the largest masses in our sample using the Python Package \texttt{WDTOOL}\footnote{\url{https://github.com/vedantchandra/wdtools}} \citep{chandra2020b}, following the procedure described in \cite{chandra2020}; which used data with the same resolution and S/N by SDSS. In brief, we fitted the line core of the H$_{\alpha}$ absorption line with three different windows around the line centroid (20 \text{\AA }, 15 \text{\AA }, and 10 \text{\AA}) as can be seen in the left panel of Figure~\ref{fig:10}. Choosing $\lambda_{\text{obs}}$ from the fit for the 10\text{ \AA } window as the best fit, the rest wavelength of H$_{\alpha}$ absorption line ($\lambda_0$) and the speed of light c, we calculated  V$_{\text{app}}$ with
\begin{equation}
    V_{\text{app}} = \left( \frac{\lambda_{\text{obs}} - \lambda_0}{\lambda_0}\right) \cdot c .
\end{equation}
The error in V$_{\text{app}}$ was obtained considering two factors, the standard deviation between the three centroid measurements for the 20 \AA, 15 \AA, and 10 \AA, and the contribution from the uncertainties of the fitting routine reported in the covariance matrix; both added in quadrature gives the error of this quantity \citep{chandra2020}. 

\begin{figure}[t!]
	\includegraphics[width=0.47\textwidth]{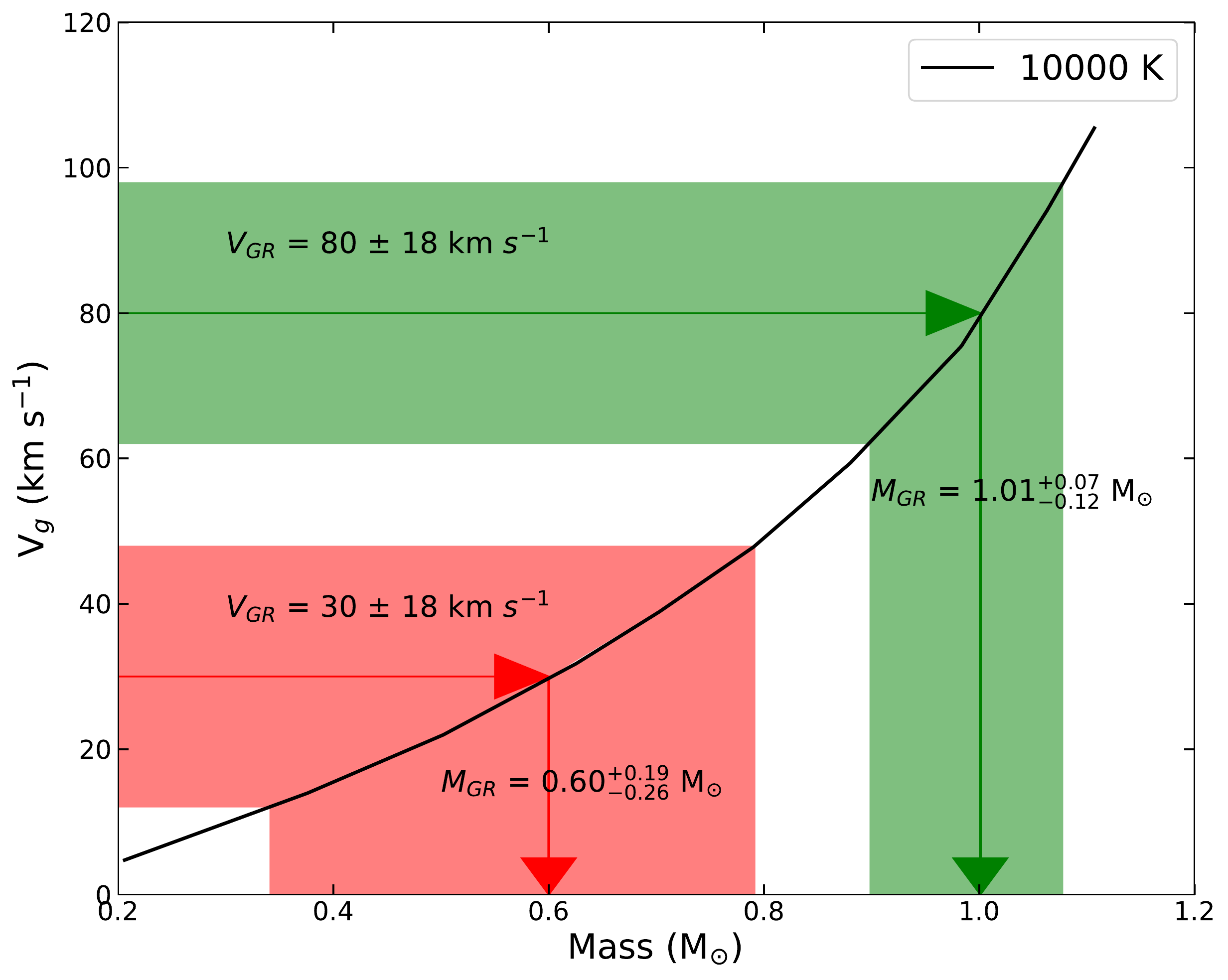}
    \caption{An example of how the velocities associated to the gravitational redshift (V$_{\text{g}}$) are converted into WD masses. The red and green arrows indicate the V$_{\text{g}}$ mapped into WD masses as obtained using the 10000 K mass-radius relation by \cite{romero2019} (black line). The red and green area indicate the errors for each measurement.}
   \label{fig:11}
  \smallskip
\end{figure}

After this, we used the heliocentric radial velocities of the TO/SG companions (i.e., V$_{r}$ in Equation~\ref{eq:11}) and the thick envelope mass-radius relation by \cite{romero2019} to obtain V$_{\text{GR}}$ and estimate the M$_{\text{GR}}$. We show an illustration of this procedure in Figure~\ref{fig:11}, for two very different values of V$_{\text{GR}}$, but assuming the same error. We can see that the mass-radius relation is less sensitive to the errors in V$_{\text{GR}}$ for the more massive WDs than it is for less massive WDs. All the parameters needed to determine M$_{\text{GR}}$ for the three WDs are listed in Table~\ref{tab:2}.
 
For the binary composed by SG3 and WD3, we obtained a WD M$_{\text{GR}}$ around 0.90 M$_{\odot}$, which is a bit bigger than the spectroscopic masses but consistent within the errors. For SG7 and WD7, we obtained a WD M$_{\text{GR}}\sim$ 0.66 M$_{\odot}$, which is in agreement with the spectroscopic measurements in this work. However, the limited resolution of the SDSS spectra available for these objects does not allow us to refine the mass determinations based on atmospheric parameters, and therefore a better determination of the masses of these interesting WDs will have to wait for better data.

Moreover, given the need of using the systemic radial velocity of the WB system in order to extract a measurement of the gravitational redshift of the WD components (V$_{\text{GR}}$ in equation~\ref{eq:11}), we had to examine to a closer level the V$_{\text{r}}$ of the primaries.  Although we derive V$_{\text{r}}$ from our own spectra, a comparison was made with determinations available in the literature. These data are displayed in the right-hand panels of Figure~\ref{fig:10}, for the two pairs under scrutiny.

Given the non negligible difference in the V$_{\text{r}}$ with time for SG3 ($\sim$ 7.00 km s$^{-1}$ between \textit{Gaia} and this work), there is a possibility of an unknown, unresolved close companion to this SG primary. More V$_{\text{r}}$ epochs are necessary to confirm this suggestion. For SG7, the difference in V$_{\text{r}}$ is below 1.00 km s$^{-1}$, not providing variability suspicions as serious as the previous case. For the rest of our sample the same analysis was done, finding no evidence of variability.

\subsubsection{Unresolved Double Degenerate Systems}
\label{sec:5.3.2}

Spectroscopic WD masses can be overestimated when dealing with an unresolved double degenerate system. These types of binaries are hard to identify when relying only on low-resolution spectra, therefore they are treated like single stars, and their characterization, both spectroscopically and photometrically, appears totally normal \citep[e.g.,][]{gb2019}.

Because these apparent single WDs are overluminous when doing the photometric analysis, the radius is overestimated and hence the mass underestimated, yielding values around $\sim$ 0.2-0.3 M$_{\odot}$. On the other hand, the spectroscopic determinations yield ``normal'' surface gravity values, thus typical WD masses \citep[e.g.,][]{bedard2017,kilic2021a}.

For our two outliers, the photometric mass was obtained from Montreal WD Database \citep{dufour2017}. For WD3, the photometric mass is M$_{\text{f}}$= 0.39 M$_{\odot}$, in comparison with the spectroscopic masses of M$_{\text{AN17}}=$ 0.70 M$_{\odot}$ and M$_{\text{KL13,KE1516}}=$ 0.82 M$_{\odot}$, with  $\Delta M_{\text{f}} \sim$ 0.30-0.40 M$_{\odot}$. For WD7, the photometric mass is M$_\text{f}$= 0.39 M$_{\odot}$, in contrast with the spectroscopic masses of M$_{\text{AN17}}=$ 0.75 M$_{\odot}$ and M$_{\text{KL13,KE1516}}=$ 0.62 M$_{\odot}$, with $\Delta M_{\text{f}} \sim$ 0.20-0.35 M$_{\odot}$. Considering this evidence, we strongly believe that these WDs are in reality double degenerate systems, and should not be used to constrain the IFMR. Both of these WDs are represented as blue stars in Figure~\ref{fig:9}.

\subsection{Metallicity dependence of the IFMR} 
\label{sec:5.4}

The semi-empirical IFMRs by AN15 and BA18 in Figure~\ref{fig:9} (black dot-dashed and solid lines, respectively) assume solar metallicities, and the impact of this parameter on theoretical IFMRs seem to be almost negligible below 2 M$_{\odot}$ \citep[e.g.,][]{dominguez1999,choi2016}. For our sample of WDs, their progenitor stars have metallicities between -0.45 to 0.50 dex, a range that is already larger than everything previously done, either from OCs or WBs, as can be seen in Table~\ref{tab:3}. The fact that this is the case just from 14 objects highlights the potential of these particular binaries for improving our understanding of metallicity effects on the IFMR.

\begin{figure}[t!]
\centering
	\includegraphics[width=0.47\textwidth]{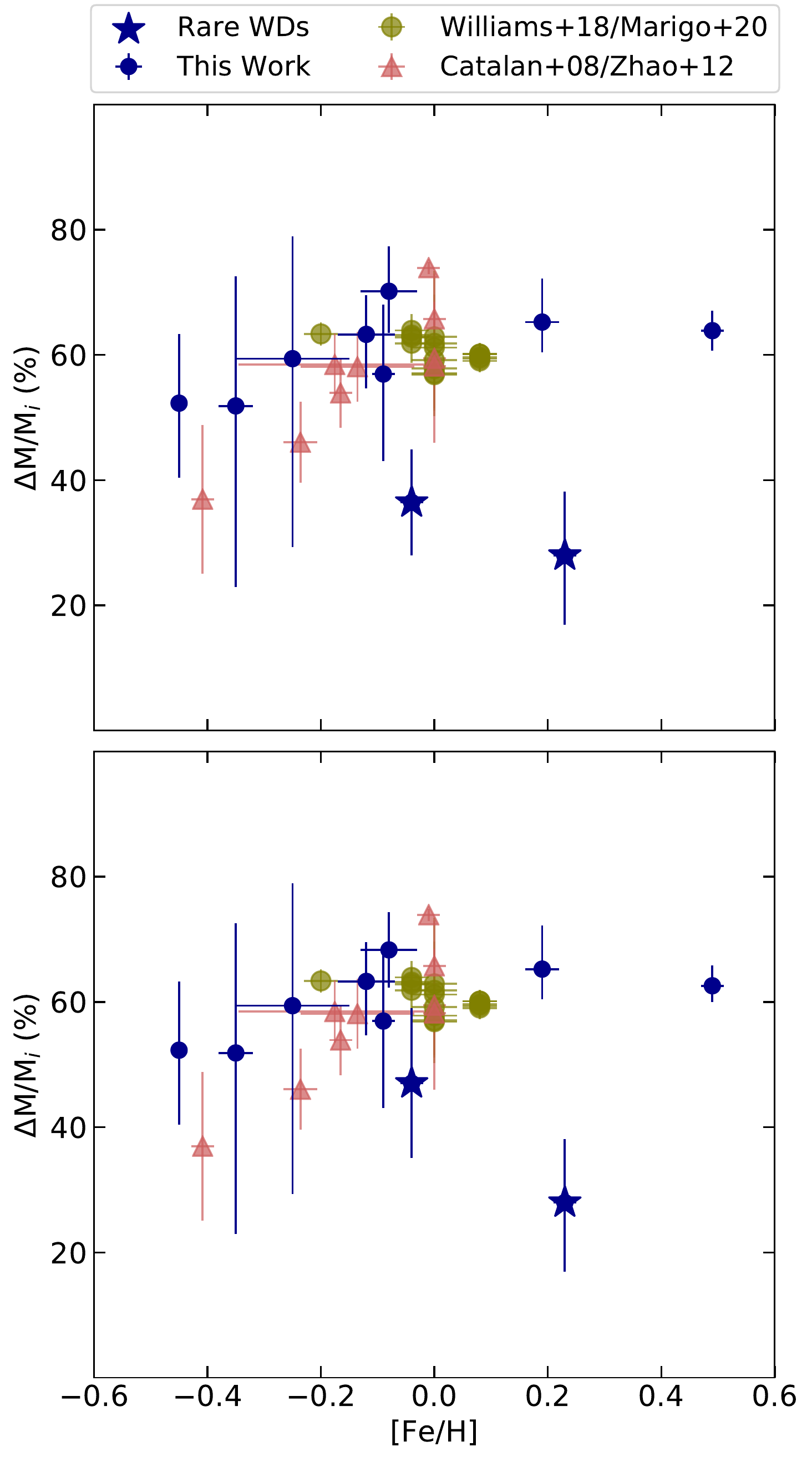}
    \caption{Semi-empirical constraints on the fraction of mass lost as a function of metallicity, for initial (progenitor) masses M$_{\text{i}} <$ 2.0 M$_{\odot}$. Red triangles are WDs from \cite{catalan2008a} and \cite{zhao2012}, green dots are WDs in OCs from \cite{williams2018} and \cite{marigo2020}, blue circles are the WDs in the present work with TO/SG primaries, and blue stars are the possible double degenerate systems. The upper panel corresponds to stellar parameters from AN17, and the lower panel to KL13 and KE1516; both of them show WDs with photometric parameters from GF19.}
   \label{fig:12}
\end{figure}

We attempt to explore the influence of metallicity on the IFMR with current constraints in Figure~\ref{fig:12}. ZH12 suggested that for a given initial mass below 2 M$_{\odot}$, metal-rich progenitors create less massive WDs than metal-poor progenitors (red triangles). For our stars (blue dots) and not considering our outliers (blue stars), we can see a similar trend for metallicities between -0.4 and 0 dex, but more points are necessary to confirm this suggestion. For [Fe/H]$\gtrsim$ -0.1 dex, the mass loss percentage seem to be constant, yielding WDs with $\sim$40\% of their progenitors masses for M$_{\text{i}}<$ 2.0 $M_{\odot}$; with the outliers showing a different behaviour ($\Delta M/M_i <$ 40\%). Overall, the metallicity does not explain the outliers nor the spread in our results, and no firmer conclusions can be extracted due to the small sample size.  Nevertheless, our results firmly establish WBs as the only available tool for empirically constraining the potential metallicity dependence of the IFMR, so expanding the sample of WBs with evolved primaries would be important to make progress in this aspect.

\begin{table}[t!]
\centering
\caption{Metallicity Range in Semi-empirical Constraints for the IFMR}
\label{tab:3}
\begin{tabular}{c c c}
\hline\hline
Constraints & [Fe/H] & Ref.\\
\hline\hline
 OCs  & -0.20 to +0.14 & WI18 and MA20\\
 MS-WD  & $\sim$ 0 & CA08\\
 MS-WD  & -0.40 to +0.19 & ZH12 \\
 SG-WD  & -0.45 to +0.50 & This Work \\
\hline\hline
\smallskip
\end{tabular}
\end{table}

\section{Summary and Conclusions}
\label{sec:6}

In this work, we have performed a search for wide binaries containing slightly evolved, pre-red giant branch primaries and white dwarf secondaries of the DA type, in order to use them to obtain semi-empirical constraints for the IFMR of WDs. The sample was selected by matching position, proper motions, and parallaxes of \textit{Gaia} surveys and SDSS DR12 WD catalog. We also selected pairs from \cite{elbadry2018} catalog with the appropriate criteria.

For the WDs, we recompiled spectroscopic atmospheric parameters existing in the literature. For the primaries, we obtained high-resolution spectra and measured stellar parameters (T$_{\text{eff}}$, log g, [Fe/H]). We also searched for stellar parameters in different mid and high resolution catalogs; which then were used to determine ages via theoretical isochrones. Our ages are insensitive to the set of isochrones used in the fitting procedure, and are in excellent agreement with seismic ages for stars in similar evolutionary stages.

Of the 15 original primaries, 13 can be classified as true turnoff/subgiant stars, for which ages with precisions better than 10\% could be obtained. The ages for these primaries range between 1.6 to 10 Gyr. Assigning these ages to the 13 WDs in their corresponding systems, we determined initial masses for their progenitors.  When paired to the independently determined WD masses, this procedure effectively provides individual semi-empirical constraints on the IFMR, one constraint per wide binary system.  All our 13 progenitor masses fall below 3.0 M$_{\odot}$, a range that has been weakly constrained previously by WDs in WBs, and poorly covered by WDs in stellar clusters.

As expected given the high precision in ages that can be achieved for slightly evolved stars, our constraints show a significant improvement in the initial mass determination when compared with those obtained before from wide binaries \citep{catalan2008a,zhao2012}.

On the other hand, the available spectroscopic WD (final) masses of our sample have large uncertainties mostly due to the low S/N of the SDSS spectra they depend on; the photometric WD masses available also show a large scatter given the broad bandpasses from \textit{Gaia} and large errors in parallax for some objects. Considering this, we can say that the precision of empirical constraints on the IFMR in the low-mass regime are now limited by the WD mass (M$_{\text{f}}$) determinations. With the use of WBs containing a WD and a TO/SG primary, we are not limited anymore by the precision in M$_{\text{i}}$, but now the pressure is on the measurement of the WD masses today.

Although still with large uncertainties in the final WD masses, if taken at face value our results would indicate a sizeable dispersion in the IFMR at the low-mass regime, larger than that observed at higher progenitor  masses from WDs in stellar clusters. This behavior is not reproduced by theoretical nor semi-empirical IFMRs, and cannot be explained by the range of metallicities spanned by our targets.  If confirmed real, such dispersion would have implications for a number of areas in astrophysics that require assuming an IFMR as one of their ingredients.

Given its potential impact, we performed a closer inspection of two of the most massive WDs in our sample. We could not confirm a radial velocity variability for the two corresponding primaries because of the lack of radial velocity epochs. However, we did notice inconsistencies between the photometric and spectroscopic solutions, finding differences in final masses $\Delta M_\text{f} \sim$ 0.20-0.40 M$_{\odot}$. These would suggest that our ``single'' WDs are in reality unresolved double degenerate systems, and therefore, they should not be used to constrain the IFMR.

Finally, we attempted our own determination of the masses of these apparently massive WDs by taking advantage of the measurement of their gravitational redshifts. However, our results were inconclusive due to the inadequate spectral resolution of the available SDSS spectra for the purposes of precisely determining the center of the narrow core of the Balmer lines.  Nevertheless, WD masses determined from gravitational redshifts should be considered a promising tool to better constrain the IFMR using wide binaries. Given the precision in initial progenitor masses as found in the present work, the motivation is now in place for devoting enough telescope time to obtain high signal-to-noise, high resolution spectra of WDs in wide binaries with TO/SG primaries, not only to finally achieve a better overall IFMR, but also to finally explore its dependence on metallicity.

\section*{Acknowledgements}
We thank the support from Chilean Time Allocation Committee (CNTAC) through program ID CN2017B-85. We thank the anonymous referee for his/her valuable comments, which helped improve our results and the paper. MB acknowledges the invaluable discussions, feedback, and support by Claudia Aguilera-Gomez, Jeff J. Andrews, Marcel Ag\"{u}eros, and Mukremin Kilic. JC acknowledges support from CONICYT project Basal AFB-170002, and from the Agencia Nacional de Investigación y Desarrollo (ANID), via Proyecto FONDECYT Regular 1191366.

\begin{appendices}

\section{Appendix: Data Tables}
\counterwithin{table}{section}
\setcounter{table}{0}

\begin{table*}[t!]
\centering
\def\arraystretch{1.5}
\setlength{\tabcolsep}{6pt}
\caption{\textbf{Wide Binary Sample}}
\label{tab:A.1}
\begin{tabular}{c c c c c c c c}
\hline\hline
Gaia Source ID &  $\mu_{\alpha}$ & $\mu_{\beta}$ & $\varpi$ &  G & $\Delta \theta$ & $\Delta V_{\perp}$ & s  \\
(DR2) & (mas yr$^{-1}$) & (mas yr$^{-1}$) & (mas) & (mag) & (arcsec) & (km s$^{-1}$) & (au) \\
\hline\hline
599615715567025024 & 11.70 $\pm$ 0.03 & -33.03 $\pm$ 0.02 & 3.70 $\pm$ 0.03 & 10.50 & 21.33 & 0.51 & 5770.55\\ 
  599615711270729984 & 12.09 $\pm$ 0.15 & -33.09 $\pm$ 0.11 & 4.06 $\pm$ 0.16 & 18.05 & & & \\ \hline 
611167253447201280 & -76.71 $\pm$ 0.02 & -0.73 $\pm$ 0.01 & 5.36 $\pm$ 0.02 & 11.60 & 19.45 & 0.19 & 3631.23\\ 
  611167150367477888 & -76.83 $\pm$ 0.09 & -0.91 $\pm$ 0.07 & 5.35 $\pm$ 0.09 & 16.99 & & & \\ \hline 
649472486214811776 & -17.15 $\pm$ 0.03 & -2.72 $\pm$ 0.02 & 3.64 $\pm$ 0.02 & 10.48 & 66.30 & 0.34 & 18230.03\\ 
  649472245695986816 & -16.99 $\pm$ 0.12 & -2.52 $\pm$ 0.07 & 3.53 $\pm$ 0.10 & 17.44 & & & \\ \hline 
722129653388999552 & -11.51 $\pm$ 0.02 & 6.15 $\pm$ 0.01 & 3.34 $\pm$ 0.02 & 11.56 & 83.59 & 0.24 & 25046.35\\ 
  722128828755291392 & -11.46 $\pm$ 0.11 & 6.00 $\pm$ 0.09 & 3.25 $\pm$ 0.12 & 17.70 & & & \\ \hline 
754878680236803328 & 5.80 $\pm$ 0.03 & 39.92 $\pm$ 0.02 & 7.50 $\pm$ 0.03 & 9.56 & 171.95 & 0.46 & 22937.65\\ 
  754878504143019136 & 5.71 $\pm$ 0.18 & 39.19 $\pm$ 0.18 & 7.84 $\pm$ 0.18 & 18.48 & & & \\ \hline 
1011166113298929664 & -47.53 $\pm$ 0.02 & -11.18 $\pm$ 0.02 & 6.24 $\pm$ 0.02 & 10.38 & 63.23 & 0.06 & 10127.94\\ 
  1011119104880885376 & -47.51 $\pm$ 0.24 & -11.11 $\pm$ 0.22 & 6.11 $\pm$ 0.26 & 19.22 & & & \\ \hline 
1478484737227019520 & 1.45 $\pm$ 0.01 & 0.98 $\pm$ 0.02 & 7.77 $\pm$ 0.02 & 11.21 & 79.08 & 0.04 & 10177.53\\ 
  1477734110083441664 & 1.39 $\pm$ 0.09 & 0.96 $\pm$ 0.12 & 7.81 $\pm$ 0.11 & 17.98 & & & \\ \hline 
1523329865151542784 & -10.46 $\pm$ 0.01 & -0.07 $\pm$ 0.01 & 3.97 $\pm$ 0.01 & 12.09 & 14.59 & 1.54 & 3673.14\\ 
  1523329865152158208 & -11.75 $\pm$ 0.08 & 0.03 $\pm$ 0.07 & 4.02 $\pm$ 0.11 & 17.91 & & & \\ \hline 
2544285956968663040 & 30.90 $\pm$ 0.02 & 22.04 $\pm$ 0.02 & 3.96 $\pm$ 0.02 & 12.01 & 11.02 & 1.39 & 2785.20\\ 
  2544285918312054784 & 31.43 $\pm$ 0.38 & 21.01 $\pm$ 0.35 & 3.86 $\pm$ 0.37 & 19.35 & & & \\ \hline 
3244617448738051072 & 35.90 $\pm$ 0.03 & 15.01 $\pm$ 0.02 & 4.69 $\pm$ 0.02 & 11.54 & 18.42 & 0.06 & 3925.73\\ 
  3244617478801862656 & 35.86 $\pm$ 0.15 & 15.04 $\pm$ 0.12 & 4.67 $\pm$ 0.14 & 17.88 & & & \\ \hline 
3637279439295576064 & -30.10 $\pm$ 0.02 & 0.68 $\pm$ 0.01 & 5.31 $\pm$ 0.02 & 8.58 & 95.70 & 0.36 & 18029.94\\ 
  3637279538079751168 & -30.50 $\pm$ 0.17 & 0.62 $\pm$ 0.09 & 5.43 $\pm$ 0.13 & 18.19 & & & \\ \hline 
3803142829929703424 & -17.55 $\pm$ 0.02 & -10.89 $\pm$ 0.02 & 5.19 $\pm$ 0.02 & 9.97 & 14.29 & 0.15 & 2755.48\\ 
  3803142859993966080 & -17.60 $\pm$ 0.13 & -10.73 $\pm$ 0.11 & 4.94 $\pm$ 0.12 & 17.86 & & & \\ \hline 
3804738874137073664 & -35.69 $\pm$ 0.02 & -84.48 $\pm$ 0.02 & 6.44 $\pm$ 0.02 & 11.35 & 62.12 & 0.12 & 9653.88\\ 
  3804750586512019968 & -35.52 $\pm$ 0.13 & -84.51 $\pm$ 0.10 & 6.66 $\pm$ 0.12 & 17.61 & & & \\ \hline 
3897577272299565056 & -29.60 $\pm$ 0.03 & -22.90 $\pm$ 0.02 & 4.46 $\pm$ 0.02 & 11.45 & 13.48 & 0.32 & 3018.53\\ 
  3897577164925150208 & -29.74 $\pm$ 0.16 & -23.17 $\pm$ 0.10 & 4.71 $\pm$ 0.16 & 17.92 & & & \\ \hline 
3930951813889966080 & 25.07 $\pm$ 0.03 & -59.39 $\pm$ 0.03 & 8.05 $\pm$ 0.03 & 11.43 & 22.32 & 0.54 & 2771.64\\ 
  3930951809593631232 & 25.15 $\pm$ 0.33 & -60.31 $\pm$ 0.27 & 8.17 $\pm$ 0.24 & 19.08 & & & \\ \hline 
736350496264030976 & 1.48 $\pm$ 0.02 & -10.76 $\pm$ 0.02 & 7.83 $\pm$ 0.02 & 9.13 & 28.64 & 0.27 & 3659.24\\ 
  736350397480224384 & 1.20 $\pm$ 0.20 & -11.11 $\pm$ 0.23 & 7.25 $\pm$ 0.23 & 18.76 & & & \\ \hline 
861577620763026048 & -21.31 $\pm$ 0.01 & -10.64 $\pm$ 0.01 & 6.72 $\pm$ 0.01 & 11.85 & 13.55 & 0.26 & 2016.23\\ 
  861577625059010176 & -21.19 $\pm$ 0.04 & -10.29 $\pm$ 0.05 & 6.72 $\pm$ 0.05 & 16.84 & & & \\ \hline 
1211355294880531968 & -61.52 $\pm$ 0.02 & 129.56 $\pm$ 0.02 & 11.64 $\pm$ 0.01 & 10.90 & 11.14 & 0.50 & 956.92\\ 
  1211355290583415296 & -62.35 $\pm$ 0.20 & 128.66 $\pm$ 0.19 & 11.65 $\pm$ 0.17 & 18.70 & & & \\ \hline 
1266120732108636160 & 14.35 $\pm$ 0.01 & -62.57 $\pm$ 0.01 & 4.49 $\pm$ 0.02 & 12.16 & 14.21 & 1.17 & 3162.42\\ 
  1266120732108636160 & 14.56 $\pm$ 0.06 & -63.66 $\pm$ 0.07 & 4.73 $\pm$ 0.08 & 17.39 & & & \\ \hline 
1274417234535214592 & -5.24 $\pm$ 0.01 & 23.89 $\pm$ 0.02 & 4.95 $\pm$ 0.02 & 11.19 & 23.82 & 0.56 & 4815.93\\ 
  1274417200175476736 & -5.68 $\pm$ 0.14 & 23.51 $\pm$ 0.18 & 4.69 $\pm$ 0.16 & 18.91 & & & \\ \hline 
1277405157384126464 & 2.38 $\pm$ 0.01 & 85.06 $\pm$ 0.02 & 8.78 $\pm$ 0.02 & 10.27 & 293.28 & 0.07 & 33409.81\\ 
  1277028742153245952 & 2.39 $\pm$ 0.07 & 85.18 $\pm$ 0.10 & 8.65 $\pm$ 0.09 & 17.93 & & & \\ \hline 
1323408242853122560 & -2.41 $\pm$ 0.01 & -75.49 $\pm$ 0.01 & 10.01 $\pm$ 0.01 & 9.80 & 130.57 & 0.07 & 13041.72\\ 
  1323407624377828352 & -2.50 $\pm$ 0.10 & -75.60 $\pm$ 0.13 & 9.93 $\pm$ 0.11 & 18.19 & & & \\ \hline 
1463032750563461120 & -47.76 $\pm$ 0.01 & 26.49 $\pm$ 0.01 & 7.17 $\pm$ 0.02 & 11.87 & 16.11 & 0.71 & 2246.31\\ 
  1463032819282938112 & -48.30 $\pm$ 0.09 & 27.43 $\pm$ 0.08 & 7.16 $\pm$ 0.11 & 17.79 & & & \\ \hline 
1519982642517952000 & 19.84 $\pm$ 0.01 & -7.85 $\pm$ 0.02 & 2.71 $\pm$ 0.02 & 12.05 & 30.88 & 1.06 & 11399.15\\ 
  1519982706941974528 & 20.06 $\pm$ 0.36 & -8.41 $\pm$ 0.51 & 2.21 $\pm$ 0.54 & 20.32 & & & \\ 
\hline\hline
\end{tabular}
\end{table*}

\begin{table*}[t!]
\centering
\def\arraystretch{1.5}
\setlength{\tabcolsep}{6pt}
    \addtocounter{table}{-1}
    \caption{\textbf{Wide Binary Sample (Cont.)}}
\label{tab:A.1c}
\begin{tabular}{c c c c c c c c}
\hline\hline
Gaia Source ID &  $\mu_{\alpha}$ & $\mu_{\beta}$ & $\varpi$ &  G & $\Delta \theta$ & $\Delta V_{\perp}$ & s  \\
(DR2) & (mas yr$^{-1}$) & (mas yr$^{-1}$) & (mas) & (mag) & (arcsec) & (km s$^{-1}$) & (au) \\
\hline\hline
2608249015120879104 & 55.67 $\pm$ 0.03 & -46.75 $\pm$ 0.02 & 17.70 $\pm$ 0.02 & 9.99 & 60.38 & 0.47 & 3410.67\\ 
  2608247533357159424 & 54.05 $\pm$ 0.10 & -47.43 $\pm$ 0.09 & 17.66 $\pm$ 0.09 & 16.94 & & & \\ \hline 
2751887496187636224 & -2.49 $\pm$ 0.03 & -8.76 $\pm$ 0.02 & 3.60 $\pm$ 0.02 & 9.35 & 38.52 & 0.72 & 10691.05\\ 
  2751887427468160000 & -1.95 $\pm$ 0.21 & -8.81 $\pm$ 0.16 & 3.61 $\pm$ 0.20 & 18.44 & & & \\ \hline 
2854365725107598336 & 1.45 $\pm$ 0.02 & -1.52 $\pm$ 0.02 & 5.29 $\pm$ 0.02 & 11.85 & 15.53 & 0.17 & 2935.62\\ 
  2854365720811907072 & 1.56 $\pm$ 0.11 & -1.68 $\pm$ 0.07 & 5.30 $\pm$ 0.11 & 17.55 & & & \\ \hline 
4032775077476713984 & 33.97 $\pm$ 0.01 & -22.71 $\pm$ 0.01 & 5.78 $\pm$ 0.01 & 10.72 & 89.00 & 0.25 & 15386.72\\ 
  4032775038821332480 & 34.00 $\pm$ 0.17 & -23.02 $\pm$ 0.16 & 5.81 $\pm$ 0.22 & 18.65 & & & \\ \hline 
4564212134294495232 & -26.88 $\pm$ 0.02 & -10.45 $\pm$ 0.02 & 9.84 $\pm$ 0.02 & 8.35 & 69.45 & 0.52 & 7061.56\\ 
  4564212061278453248 & -27.40 $\pm$ 0.17 & -11.39 $\pm$ 0.17 & 9.80 $\pm$ 0.20 & 18.86 & & & \\ \hline 
6908406487039810560 & -6.36 $\pm$ 0.03 & -4.92 $\pm$ 0.02 & 6.79 $\pm$ 0.03 & 11.07 & 236.06 & 0.39 & 34756.23\\ 
  6908407066861225984 & -6.61 $\pm$ 0.26 & -5.43 $\pm$ 0.21 & 7.10 $\pm$ 0.25 & 18.86 & & & \\ 
\hline\hline
\end{tabular}
\end{table*}

\begin{table*}[!htbp]
\centering
    \caption{\textbf{Physical Properties of White Dwarfs in our Wide Binaries}}
    \label{tab:A.2}
   \def\arraystretch{1.7}
   \setlength{\tabcolsep}{10pt}
\resizebox{\textwidth}{!}{
    \begin{tabular}{c c c c c c c c c c c c}
    \hline\hline
Pair N\textdegree & Name & Gaia source ID & S/N & T$_{\text{eff}}$ & log(g) &  M$_{\text{f}}$ & t$_{\text{cool}}$ & t$_{\text{prog}}$ & M$_{\text{i}}$ & Ref. \\
 & & (DR2) & & (K) & (cm s$^{-2}$) & (M$_{\odot}$) & (Gyr) & (Gyr) & (M$_{\odot}$) &  \\
\hline\hline 
\multicolumn{10}{c}{Spectroscopic Parameters}\\
\hline\hline
1 & WD1 & 2751887427468160000 & 26 & 16717 $\pm$ 304 & 7.85 $\pm$ 0.08 & 0.54 $^{+0.04}_{-0.04}$ & 0.10 $^{+0.02}_{-0.01}$ & 1.58 $_{-0.16}^{+0.18}$ & 1.81 $_{-0.08}^{+0.09}$  & 1 \\
 & & &  & 16451 $\pm$ 230 & 7.94 $\pm$ 0.04 & 0.58 $^{+0.03}_{-0.02}$ & 0.13 $^{+0.01}_{-0.02}$ &  1.55 $_{-0.16}^{+0.18}$ & 1.83 $_{-0.08}^{+0.09}$ & 3 \\
\hline
2 & WD2 & 593918802224747776 & 20 & 15782 $\pm$ 368 & 7.98 $\pm$ 0.08 & 0.60 $^{+0.05}_{-0.04}$ & 0.17 $^{+0.03}_{-0.02}$ &  0.23 $_{-0.18}^{+0.86}$ & 3.88 $_{-1.61}^{+1.48}$  & 1 \\
 & & & & 16143 $\pm$ 226 & 8.19 $\pm$ 0.04 & 0.73 $^{+0.03}_{-0.02}$ & 0.22 $^{+0.02}_{-0.01}$ &  0.18 $_{-0.18}^{+0.86}$ & 4.26 $_{-1.95}^{+1.15}$  & 2 \\
\hline
3 & WD3 & 668562516332550784 & 30 & 9166 $\pm$ 128 & 8.17 $\pm$ 0.07 & 0.70 $^{+0.05}_{-0.04}$ & 1.01 $^{+0.12}_{-0.09}$ &  1.33 $_{-0.25}^{+0.21}$ & 2.04 $_{-0.13}^{+0.21}$ & 1 \\
 & & &  & 9175 $\pm$ 128 & 8.35 $\pm$ 0.04 & 0.82 $^{+0.03}_{-0.03}$ & 1.34 $^{+0.12}_{-0.09}$ & 1.00 $_{-0.25}^{+0.21}$ & 2.32 $_{-0.17}^{+0.24}$ & 2 \\
\hline
4 & WD4 & 3244617478801862656 & 41 & 16149 $\pm$ 226 & 7.85 $\pm$ 0.05 & 0.53 $^{+0.03}_{-0.02}$ & 0.12 $^{+0.01}_{-0.01}$ &  0.38 $_{-0.30}^{+5.40}$ & 3.05 $_{-1.90}^{+2.66}$  & 1 \\
 & & & & 15949 $\pm$ 223 & 7.91 $\pm$ 0.04 & 0.57 $^{+0.02}_{-0.02}$ & 0.14 $^{+0.01}_{-0.01}$ & 0.36 $_{-0.30}^{+5.40}$ & 3.11 $_{-1.96}^{+3.39}$ & 2\\
\hline
5 & WD5 & 3637279538079751168 & 19 & 13019 $\pm$ 371 & 8.23 $\pm$ 0.11 & 0.75 $^{+0.07}_{-0.07}$ & 0.44 $^{+0.08}_{-0.07}$ & - & - & 1 \\  
 & & & & 13218 $\pm$ 215 & 8.30 $\pm$ 0.07 & 0.80 $^{+0.05}_{-0.04}$ & 0.47 $^{+0.05}_{-0.05}$ & - & - & 2\\
\hline
6 & WD6 & 3803142859993965952 & 47 & 15601 $\pm$ 218  & 7.91 $\pm$ 0.04 & 0.56 $^{+0.03}_{-0.02}$ & 0.15 $^{+0.01}_{-0.01}$ &  3.49 $_{-0.20}^{+0.20}$ & 1.55 $_{-0.03}^{+0.02}$ & 1 \\
 & & & & 15422 $\pm$ 215 & 7.94 $\pm$ 0.04 &  0.58 $^{+0.02}_{-0.02}$ & 0.16 $^{+0.01}_{-0.01}$ & 3.48 $_{-0.20}^{+0.20}$ &  1.55 $_{-0.03}^{+0.02}$ & 2 \\
\hline
7 & WD7 & 3975115240311668096 & 15 & 11883 $\pm$ 407 & 8.24 $\pm$ 0.14  & 0.75 $^{+0.09}_{-0.08}$ & 0.57 $^{+0.14}_{-0.11}$ & 6.19 $_{-0.32}^{+0.32}$ & 1.18 $_{-0.01}^{+0.02}$ & 1 \\
 & & & & 11909 $\pm$ 1309 & 8.02 $\pm$ 0.21 & 0.62 $^{+0.13}_{-0.12}$ & 0.41 $^{+0.20}_{-0.15}$ & 6.35 $_{-0.34}^{+0.35}$ & 1.17 $_{-0.01}^{+0.02}$ & 2 \\
\hline\hline
\multicolumn{10}{c}{Photometric Parameters}\\
\hline\hline
8 & WD8 & 93238895273124480 & & 6451 $\pm$ 324 & 8.05 $\pm$ 0.18 & 0.62 $_{-0.11}^{+0.12}$ & 2.07 $_{-0.25}^{+0.63}$ & 3.19 $_{-0.34}^{+0.67}$ & 1.44 $_{-0.08}^{+0.05}$ &  4\\
\hline
9 & WD9 & 2198431172852758656 & & 16467 $\pm$ 145 & 8.06 $\pm$ 0.04 & 0.65 $_{-0.02}^{+0.02}$ & 0.17 $_{-0.01}^{+0.02}$ &  1.78 $_{-0.24}^{+0.18}$ & 1.87 $_{-0.07}^{+0.11}$ & 4\\ 
\hline
10 & WD10 & 6159300693821280000 & & 6478 $\pm$ 802 & 7.92 $\pm$ 0.44 & 0.54 $_{-0.21}^{+0.28}$ & 1.70 $_{-0.34}^{+1.24}$ & 3.53 $_{-0.42}^{+1.25}$ & 1.33 $_{-0.12}^{+0.05}$ & 4 \\
\hline
11 & WD11 & 2879667033851088896 & & 6157 $\pm$ 78 & 8.21 $\pm$ 0.05 & 0.72 $_{-0.03}^{+0.03}$ & 3.59 $_{-0.31}^{+0.26}$ & 0.41 $_{-0.36}^{+0.35}$ & 3.00 $_{-0.59}^{+4.00}$ & 4 \\
\hline
12 & WD12 & 64009654254079232 & & 8856 $\pm$ 421 & 8.40 $\pm$ 0.16 & 0.85 $_{-0.10}^{+0.11}$ & 1.71 $_{-0.32}^{+0.48}$ & 7.73 $_{-0.40}^{+0.57}$ & 1.18 $_{-0.02}^{+0.02}$ & 4 \\
\hline
13 & WD13 & 129072494619153536 & & 8037 $\pm$ 433 & 7.86 $\pm$ 0.21 & 0.52 $_{-0.11}^{+0.12}$ & 0.92 $_{-0.08}^{+0.14}$ & 6.08 $_{-0.24}^{+0.24}$ & 1.09 $_{-0.01}^{+0.01}$ & 4 \\
\hline
14 & WD14 & 1999564768169857792 & & 4648 $\pm$ 87 & 7.82 $\pm$ 0.07 & 0.47 $_{-0.03}^{+0.04}$ & 5.46 $_{-0.44}^{+0.86}$ &  4.46 $_{-0.56}^{+0.88}$ & 1.28 $_{-0.07}^{+0.05}$ & 4 \\ 
\hline
15 & WD15 &2153552814748001792 & & 7480 $\pm$ 898 & 8.10 $\pm$ 0.41 & 0.65 $_{-0.23}^{+0.27}$ & 1.54 $_{-0.31}^{+1.12}$ & 3.27 $_{-0.33}^{+1.12}$ & 1.35 $_{-0.12}^{+0.05}$ & 4 \\ 
    \hline\hline
    \end{tabular}}
\begin{tablenotes}
\small
\item \textbf{References.} (1)\cite{anguiano2017}; (2)\cite{kleinman2013}; (3)\cite{kepler2015,kepler2016}; (4)\cite{gf2019}.
\end{tablenotes}
\end{table*}

\begin{table*}[!htbp]
\centering
    \caption{\textbf{Atmospheric Parameters for the Turnoff/Subgiant Companion 
    Observed in our Wide Binaries}}
    \label{tab:A.3}
    \def\arraystretch{1.8}
    \setlength{\tabcolsep}{10pt}
    \resizebox{\textwidth}{!}{
    \begin{tabular}{c c c c c c c c c c c c}
    \hline\hline
Pair N\textdegree & Name & Gaia source ID & S/N & T$_{\text{eff}}$ & log g & [Fe/H] & M$_{\text{SG}}$ & Age \texttt{YY} & Age \texttt{MIST} & Age \texttt{PARAM} & Ref.\\
&  & (DR2) &  & (K) & (cm s$^{-2}$) & (dex) & (M$_{\odot}$) & (Gyr) & (Gyr) & (Gyr) & \\
    \hline\hline
\multicolumn{11}{c}{Observed TO/SG}\\
\hline\hline
1 & SG1 & 2751887496187636096 & 123 & 6564 $\pm$ 83 & 3.90 $\pm$ 0.15 & -0.08 $\pm$ 0.05 & 1.65 $\pm$ 0.04 & 1.81 $^{+0.19}_{-0.18}$ & 1.68 $^{+0.18}_{-0.16}$ & 1.83 $\pm$ 0.14 & 1 \\
2 & SG2 & 593915842991753088 & 79 & 6668 $\pm$ 131 & 4.53 $\pm$ 0.22 & 0.11 $\pm$ 0.08 & 1.33 $\pm$ 0.05 & 0.45 $^{+0.86}_{-0.18}$ & 0.40  $^{+0.86}_{-0.18}$ & 0.63 $\pm$ 0.48 & 1\\
3 & SG3 & 668561279381974400 & 185 & 6408 $\pm$ 64 & 4.01 $\pm$ 0.10 & 0.06 $\pm$ 0.04 & 1.44 $\pm$ 0.03 & 2.46 $^{+0.19}_{-0.24}$ & 2.34 $^{+0.19}_{-0.23}$ & 2.32 $\pm$ 0.21 & 1\\
4 & SG4 & 3244617448738051200 & 166 & 5880 $\pm$ 19 & 4.45 $\pm$ 0.06 & -0.31 $\pm$ 0.02 & 1.04 $\pm$ 0.03 & 0.89 $^{+1.16}_{-0.71}$ & 0.50 $^{+5.40}_{-0.30}$ & 0.33 $\pm$ 0.22 & 1\\
6 & SG6 & 3803142829929703552 & 216 & 5798 $\pm$ 23 & 4.19 $\pm$ 0.04 & 0.49 $\pm$ 0.02 & 1.35 $\pm$ 0.03 &  3.56 $^{+0.21}_{-0.22}$ & 3.64 $^{+0.20}_{-0.20}$ & 3.42 $\pm$ 0.12 & 1\\
7 & SG7 & 3975115244607118848 & 147 & 5281 $\pm$ 19 & 3.88 $\pm$ 0.05 & -0.04 $\pm$ 0.02 & 1.12 $\pm$ 0.02 & 7.15 $^{+0.30}_{-0.28}$ & 6.76 $^{+0.29}_{-0.30}$ & 7.31 $\pm$ 0.14 & 1\\
\hline\hline
\multicolumn{11}{c}{Literature TO/SG}\\
\hline\hline
8 & SG8 & 93238895273066752 & 508 & 6180 $\pm$ 14 & 4.16 $\pm$ 0.02 & -0.09 $\pm$ 0.02 & 1.14 $\pm$ 0.02 & 3.64 $^{+0.21}_{-0.22}$ & 5.26 $^{+0.23}_{-0.23}$  & 4.64 $\pm$ 0.29 & 3\\
9 & SG9 & 2198430859305721344 & $>$100 & 6261 $\pm$ 44 & 3.78 $\pm$ 0.02 & 0.19 $\pm$ 0.03 & 1.67 $\pm$ 0.02 & 2.59 $^{+0.17}_{-0.25}$ & 1.95 $^{+0.18}_{-0.24}$ & 1.89 $\pm$ 0.10 & 2\\
10 & SG10 & 6159300796901641856 & 86 & 6258 $\pm$ 54 & 4.13 $\pm$ 0.04 & -0.25 $\pm$ 0.10 & 1.13 $\pm$ 0.02 &  4.18 $^{+0.40}_{-0.37}$ & 5.23 $^{+0.18}_{-0.24}$ & 4.76 $\pm$ 0.90 & 4\\
11 & SG11 & 2879667068210826752 & $>$100 & 6040 $\pm$ 64 & 3.83 $\pm$ 0.03 & -0.24 $\pm$ 0.06 & 1.23 $\pm$ 0.03 & 4.00 $^{+0.22}_{-0.20}$ & 4.00 $^{+0.23}_{-0.19}$ & 3.88 $\pm$ 0.21 & 2\\
12 & SG12 & 64009276298551424 & 660  & 5505 $\pm$ 26 & 4.05 $\pm$ 0.04 & 0.23 $\pm$ 0.02 & 1.07 $\pm$ 0.02 & 9.29 $^{+0.32}_{-0.26}$ & 9.44 $^{+0.30}_{-0.24}$ & 9.58 $\pm$ 0.24 & 3\\
13 & SG13 & 129073632786900480 & 560 & 5968 $\pm$ 13 & 3.92 $\pm$ 0.02 & -0.45 $\pm$ 0.01 & 1.01 $\pm$ 0.02 & 6.40 $^{+0.10}_{-0.10}$ & 7.00 $^{+0.12}_{-0.10}$ & 7.24 $\pm$ 0.10 & 3\\
14 & SG14 & 1999563943535301760 & $>$100 & 5757 $\pm$ 36 & 4.10 $\pm$ 0.03 & -0.12 $\pm$ 0.05 & 0.98 $\pm$ 0.02 & 9.80 $^{+0.16}_{-0.50}$ & 9.92 $^{+0.19}_{-0.34}$ & 9.99 $\pm$ 0.59 & 2\\
15 & SG15 & 2153552647245580544 & 98 & 6170 $\pm$ 36 & 3.89 $\pm$ 0.02 & -0.35 $\pm$ 0.03 & 1.14 $\pm$ 0.02 & 4.21 $^{+0.10}_{-0.12}$ & 4.81 $^{+0.10}_{-0.12}$ & 4.73 $\pm$ 0.14 & 3\\
\hline\hline
\end{tabular}}
\begin{tablenotes}
\small
\item \textbf{References.} (1)This work; (2)\cite{soubiran2016}; (3)\cite{lamost2019}; (4)\cite{rave2017}.
\end{tablenotes}
\end{table*}

\end{appendices}

\clearpage
\bibliographystyle{yahapj}
\bibliography{ApJ}

\begin{thebibliography}{}
\providecommand\natexlab[1]{#1}
\providecommand\JournalTitle[1]{#1}

\bibitem[{{Aguilera-G{\'o}mez} {et~al.}(2018){Aguilera-G{\'o}mez},
  {Ram{\'\i}rez}, \& {Chanam{\'e}}}]{ag2018}
{Aguilera-G{\'o}mez}, C., {Ram{\'\i}rez}, I., \& {Chanam{\'e}}, J. 2018,
  \href{http://dx.doi.org/10.1051/0004-6361/201732209}{\JournalTitle{\aap},
  614, A55}

\bibitem[{{Andrews} {et~al.}(2015){Andrews}, {Ag{\"u}eros}, {Gianninas},
  {Kilic}, {Dhital}, \& {Anderson}}]{andrews2015}
{Andrews}, J.~J., {Ag{\"u}eros}, M.~A., {Gianninas}, A., {et~al.} 2015,
  \href{http://dx.doi.org/10.1088/0004-637X/815/1/63}{\JournalTitle{\apj}, 815,
  63}

\bibitem[{{Andrews} {et~al.}(2019){Andrews}, {Anguiano}, {Chanam{\'e}},
  {Ag{\"u}eros}, {Lewis}, {Hayes}, \& {Majewski}}]{andrews2019}
{Andrews}, J.~J., {Anguiano}, B., {Chanam{\'e}}, J., {et~al.} 2019,
  \href{http://dx.doi.org/10.3847/1538-4357/aaf502}{\JournalTitle{\apj}, 871,
  42}

\bibitem[{{Andrews} {et~al.}(2017){Andrews}, {Chanam{\'e}}, \&
  {Ag{\"u}eros}}]{andrews2017}
{Andrews}, J.~J., {Chanam{\'e}}, J., \& {Ag{\"u}eros}, M.~A. 2017,
  \href{http://dx.doi.org/10.1093/mnras/stx2000}{\JournalTitle{\mnras}, 472,
  675}

\bibitem[{{Andrews} {et~al.}(2018){Andrews}, {Chanam{\'e}}, \&
  {Ag{\"u}eros}}]{andrews2018}
{Andrews}, J.~J., {Chanam{\'e}}, J., \& {Ag{\"u}eros}, M.~A. 2018,
  \href{http://dx.doi.org/10.1093/mnras/stx2685}{\JournalTitle{\mnras}, 473,
  5393}

\bibitem[{{Anguiano} {et~al.}(2017){Anguiano}, {Rebassa-Mansergas},
  {Garc{\'{\i}}a-Berro}, {Torres}, {Freeman}, \& {Zwitter}}]{anguiano2017}
{Anguiano}, B., {Rebassa-Mansergas}, A., {Garc{\'{\i}}a-Berro}, E., {et~al.}
  2017, \href{http://dx.doi.org/10.1093/mnras/stx796}{\JournalTitle{\mnras},
  469, 2102}

\bibitem[{{Bailer-Jones} {et~al.}(2018){Bailer-Jones}, {Rybizki}, {Fouesneau},
  {Mantelet}, \& {Andrae}}]{bj2018}
{Bailer-Jones}, C.~A.~L., {Rybizki}, J., {Fouesneau}, M., {Mantelet}, G., \&
  {Andrae}, R. 2018,
  \href{http://dx.doi.org/10.3847/1538-3881/aacb21}{\JournalTitle{\aj}, 156,
  58}

\bibitem[{{B{\'e}dard} {et~al.}(2020){B{\'e}dard}, {Bergeron}, {Brassard}, \&
  {Fontaine}}]{bedard2020}
{B{\'e}dard}, A., {Bergeron}, P., {Brassard}, P., \& {Fontaine}, G. 2020,
  \href{http://dx.doi.org/10.3847/1538-4357/abafbe}{\JournalTitle{\apj}, 901,
  93}

\bibitem[{{B{\'e}dard} {et~al.}(2017){B{\'e}dard}, {Bergeron}, \&
  {Fontaine}}]{bedard2017}
{B{\'e}dard}, A., {Bergeron}, P., \& {Fontaine}, G. 2017,
  \href{http://dx.doi.org/10.3847/1538-4357/aa8bb6}{\JournalTitle{\apj}, 848,
  11}

\bibitem[{{Bergeron} {et~al.}(1992){Bergeron}, {Saffer}, \&
  {Liebert}}]{bergeron1992}
{Bergeron}, P., {Saffer}, R.~A., \& {Liebert}, J. 1992,
  \href{http://dx.doi.org/10.1086/171575}{\JournalTitle{\apj}, 394, 228}

\bibitem[{{Bernstein} {et~al.}(2003){Bernstein}, {Shectman}, {Gunnels},
  {Mochnacki}, \& {Athey}}]{bernstein2003}
{Bernstein}, R., {Shectman}, S.~A., {Gunnels}, S.~M., {Mochnacki}, S., \&
  {Athey}, A.~E. 2003, \href{http://dx.doi.org/10.1117/12.461502}{in \procspie,
  Vol. 4841, Instrument Design and Performance for Optical/Infrared
  Ground-based Telescopes, ed. M.~{Iye} \& A.~F.~M. {Moorwood}}, 1694

\bibitem[{{Bressan} {et~al.}(2012){Bressan}, {Marigo}, {Girardi}, {Salasnich},
  {Dal Cero}, {Rubele}, \& {Nanni}}]{bressan2012}
{Bressan}, A., {Marigo}, P., {Girardi}, L., {et~al.} 2012,
  \href{http://dx.doi.org/10.1111/j.1365-2966.2012.21948.x}{\JournalTitle{\mnras},
  427, 127}

\bibitem[{{Bruntt} {et~al.}(2012){Bruntt}, {Basu}, {Smalley}, {Chaplin},
  {Verner}, {Bedding}, {Catala}, {Gazzano}, {Molenda-{\.Z}akowicz}, {Thygesen},
  {Uytterhoeven}, {Hekker}, {Huber}, {Karoff}, {Mathur}, {Mosser},
  {Appourchaux}, {Campante}, {Elsworth}, {Garc{\'\i}a}, {Handberg}, {Metcalfe},
  {Quirion}, {R{\'e}gulo}, {Roxburgh}, {Stello}, {Christensen-Dalsgaard},
  {Kawaler}, {Kjeldsen}, {Morris}, {Quintana}, \& {Sanderfer}}]{bruntt2012}
{Bruntt}, H., {Basu}, S., {Smalley}, B., {et~al.} 2012,
  \href{http://dx.doi.org/10.1111/j.1365-2966.2012.20686.x}{\JournalTitle{\mnras},
  423, 122}

\bibitem[{{Buchhave} \& {Latham}(2015)}]{bl2015}
{Buchhave}, L.~A. \& {Latham}, D.~W. 2015,
  \href{http://dx.doi.org/10.1088/0004-637X/808/2/187}{\JournalTitle{\apj},
  808, 187}

\bibitem[{{Canton} {et~al.}(2021){Canton}, {Williams}, {Kilic}, \&
  {Bolte}}]{canton2021}
{Canton}, P.~A., {Williams}, K.~A., {Kilic}, M., \& {Bolte}, M. 2021,
  \href{http://dx.doi.org/10.3847/1538-3881/abe1ad}{\JournalTitle{\aj}, 161,
  169}

\bibitem[{Casagrande(2020)}]{Casagrande2020}
Casagrande, L. 2020,
  \href{http://dx.doi.org/10.3847/1538-4357/ab929f}{\JournalTitle{The
  Astrophysical Journal}, 896, 26}

\bibitem[{{Casagrande} {et~al.}(2011){Casagrande}, {Sch{\"o}nrich}, {Asplund},
  {Cassisi}, {Ram{\'\i}rez}, {Mel{\'e}ndez}, {Bensby}, \&
  {Feltzing}}]{casagrande2011}
{Casagrande}, L., {Sch{\"o}nrich}, R., {Asplund}, M., {et~al.} 2011,
  \href{http://dx.doi.org/10.1051/0004-6361/201016276}{\JournalTitle{\aap},
  530, A138}

\bibitem[{{Catal{\'a}n} {et~al.}(2008){Catal{\'a}n}, {Isern},
  {Garc{\'{\i}}a-Berro}, {Ribas}, {Allende Prieto}, \&
  {Bonanos}}]{catalan2008a}
{Catal{\'a}n}, S., {Isern}, J., {Garc{\'{\i}}a-Berro}, E., {et~al.} 2008,
  \href{http://dx.doi.org/10.1051/0004-6361:20078111}{\JournalTitle{\aap}, 477,
  213}

\bibitem[{Chabrier(2001)}]{chabrier2001}
Chabrier, G. 2001, \href{http://dx.doi.org/10.1086/321401}{\JournalTitle{The
  Astrophysical Journal}, 554, 1274}

\bibitem[{{Chanam{\'e}}(2007)}]{chaname2007}
{Chanam{\'e}}, J. 2007, \href{http://dx.doi.org/10.1017/S1743921307004243}{in
  Binary Stars as Critical Tools \& Tests in Contemporary Astrophysics, ed.
  W.~I. {Hartkopf}, P.~{Harmanec}, \& E.~F. {Guinan}, Vol. 240}, 316

\bibitem[{{Chanam{\'e}} \& {Gould}(2004)}]{c&g2004}
{Chanam{\'e}}, J. \& {Gould}, A. 2004,
  \href{http://dx.doi.org/10.1086/380442}{\JournalTitle{\apj}, 601, 289}

\bibitem[{{Chanam{\'e}} \& {Ram{\'{\i}}rez}(2012)}]{cr12}
{Chanam{\'e}}, J. \& {Ram{\'{\i}}rez}, I. 2012,
  \href{http://dx.doi.org/10.1088/0004-637X/746/1/102}{\JournalTitle{\apj},
  746, 102}

\bibitem[{{Chandra} {et~al.}(2020{\natexlab{a}}){Chandra}, {Hwang}, {Zakamska},
  \& {Budav{\'a}ri}}]{chandra2020b}
{Chandra}, V., {Hwang}, H.-C., {Zakamska}, N.~L., \& {Budav{\'a}ri}, T.
  2020{\natexlab{a}},
  \href{http://dx.doi.org/10.1093/mnras/staa2165}{\JournalTitle{\mnras}, 497,
  2688}

\bibitem[{{Chandra} {et~al.}(2020{\natexlab{b}}){Chandra}, {Hwang}, {Zakamska},
  \& {Cheng}}]{chandra2020}
{Chandra}, V., {Hwang}, H.-C., {Zakamska}, N.~L., \& {Cheng}, S.
  2020{\natexlab{b}},
  \href{http://dx.doi.org/10.3847/1538-4357/aba8a2}{\JournalTitle{\apj}, 899,
  146}

\bibitem[{{Chaplin} {et~al.}(2014){Chaplin}, {Basu}, {Huber}, {Serenelli},
  {Casagrande}, {Silva Aguirre}, {Ball}, {Creevey}, {Gizon}, {Handberg},
  {Karoff}, {Lutz}, {Marques}, {Miglio}, {Stello}, {Suran}, {Pricopi},
  {Metcalfe}, {Monteiro}, {Molenda-{\.Z}akowicz}, {Appourchaux},
  {Christensen-Dalsgaard}, {Elsworth}, {Garc{\'\i}a}, {Houdek}, {Kjeldsen},
  {Bonanno}, {Campante}, {Corsaro}, {Gaulme}, {Hekker}, {Mathur}, {Mosser},
  {R{\'e}gulo}, \& {Salabert}}]{chaplin2014}
{Chaplin}, W.~J., {Basu}, S., {Huber}, D., {et~al.} 2014,
  \href{http://dx.doi.org/10.1088/0067-0049/210/1/1}{\JournalTitle{\apjs}, 210,
  1}

\bibitem[{{Choi} {et~al.}(2016){Choi}, {Dotter}, {Conroy}, {Cantiello},
  {Paxton}, \& {Johnson}}]{choi2016}
{Choi}, J., {Dotter}, A., {Conroy}, C., {et~al.} 2016,
  \href{http://dx.doi.org/10.3847/0004-637X/823/2/102}{\JournalTitle{\apj},
  823, 102}

\bibitem[{{Cummings} {et~al.}(2019){Cummings}, {Kalirai}, {Choi}, {Georgy},
  {Tremblay}, \& {Ramirez-Ruiz}}]{cummings2019}
{Cummings}, J.~D., {Kalirai}, J.~S., {Choi}, J., {et~al.} 2019,
  \href{http://dx.doi.org/10.3847/2041-8213/aafc2d}{\JournalTitle{\apjl}, 871,
  L18}

\bibitem[{{Cummings} {et~al.}(2018){Cummings}, {Kalirai}, {Tremblay},
  {Ramirez-Ruiz}, \& {Choi}}]{cummings2018}
{Cummings}, J.~D., {Kalirai}, J.~S., {Tremblay}, P.-E., {Ramirez-Ruiz}, E., \&
  {Choi}, J. 2018,
  \href{http://dx.doi.org/10.3847/1538-4357/aadfd6}{\JournalTitle{\apj}, 866,
  21}

\bibitem[{{da Silva} {et~al.}(2006){da Silva}, {Girardi}, {Pasquini},
  {Setiawan}, {von der L{\"u}he}, {de Medeiros}, {Hatzes}, {D{\"o}llinger}, \&
  {Weiss}}]{dasilva2006}
{da Silva}, L., {Girardi}, L., {Pasquini}, L., {et~al.} 2006,
  \href{http://dx.doi.org/10.1051/0004-6361:20065105}{\JournalTitle{\aap}, 458,
  609}

\bibitem[{{Demarque} {et~al.}(2004){Demarque}, {Woo}, {Kim}, \&
  {Yi}}]{demarque2004}
{Demarque}, P., {Woo}, J.-H., {Kim}, Y.-C., \& {Yi}, S.~K. 2004,
  \href{http://dx.doi.org/10.1086/424966}{\JournalTitle{\apjs}, 155, 667}

\bibitem[{{Dobbie} {et~al.}(2006){Dobbie}, {Napiwotzki}, {Burleigh}, {Barstow},
  {Boyce}, {Casewell}, {Jameson}, {Hubeny}, \& {Fontaine}}]{dobbie2006}
{Dobbie}, P.~D., {Napiwotzki}, R., {Burleigh}, M.~R., {et~al.} 2006,
  \href{http://dx.doi.org/10.1111/j.1365-2966.2006.10311.x}{\JournalTitle{\mnras},
  369, 383}

\bibitem[{{Dominguez} {et~al.}(1999){Dominguez}, {Chieffi}, {Limongi}, \&
  {Straniero}}]{dominguez1999}
{Dominguez}, I., {Chieffi}, A., {Limongi}, M., \& {Straniero}, O. 1999,
  \href{http://dx.doi.org/10.1086/307787}{\JournalTitle{\apj}, 524, 226}

\bibitem[{{Dotter}(2016)}]{dotter2016}
{Dotter}, A. 2016,
  \href{http://dx.doi.org/10.3847/0067-0049/222/1/8}{\JournalTitle{\apjs}, 222,
  8}

\bibitem[{{Dufour} {et~al.}(2017){Dufour}, {Blouin}, {Coutu},
  {Fortin-Archambault}, {Thibeault}, {Bergeron}, \& {Fontaine}}]{dufour2017}
{Dufour}, P., {Blouin}, S., {Coutu}, S., {et~al.} 2017, in Astronomical Society
  of the Pacific Conference Series, Vol. 509, 20th European White Dwarf
  Workshop, ed. P.-E. {Tremblay}, B.~{Gaensicke}, \& T.~{Marsh}, 3

\bibitem[{{Eisenstein} {et~al.}(2006){Eisenstein}, {Liebert}, {Harris},
  {Kleinman}, {Nitta}, {Silvestri}, {Anderson}, {Barentine}, {Brewington},
  {Brinkmann}, {Harvanek}, {Krzesi{\'n}ski}, {Neilsen}, {Long}, {Schneider}, \&
  {Snedden}}]{eisenstein2006}
{Eisenstein}, D.~J., {Liebert}, J., {Harris}, H.~C., {et~al.} 2006,
  \href{http://dx.doi.org/10.1086/507110}{\JournalTitle{\apjs}, 167, 40}

\bibitem[{{El-Badry} \& {Rix}(2018)}]{elbadry2018}
{El-Badry}, K. \& {Rix}, H.-W. 2018,
  \href{http://dx.doi.org/10.1093/mnras/sty2186}{\JournalTitle{\mnras}, 480,
  4884}

\bibitem[{{El-Badry} {et~al.}(2021){El-Badry}, {Rix}, \&
  {Heintz}}]{elbadry2021}
{El-Badry}, K., {Rix}, H.-W., \& {Heintz}, T.~M. 2021,
  \href{http://dx.doi.org/10.1093/mnras/stab323}{\JournalTitle{\mnras}, 506,
  2269}

\bibitem[{{El-Badry} {et~al.}(2018){El-Badry}, {Rix}, \&
  {Weisz}}]{elbadryb2018}
{El-Badry}, K., {Rix}, H.-W., \& {Weisz}, D.~R. 2018,
  \href{http://dx.doi.org/10.3847/2041-8213/aaca9c}{\JournalTitle{\apjl}, 860,
  L17}

\bibitem[{{Espinoza-Rojas} {et~al.}(2021){Espinoza-Rojas}, {Chanam{\'e}},
  {Jofr{\'e}}, \& {Casamiquela}}]{fran2021}
{Espinoza-Rojas}, F., {Chanam{\'e}}, J., {Jofr{\'e}}, P., \& {Casamiquela}, L.
  2021, \JournalTitle{arXiv e-prints}, arXiv:2105.01096

\bibitem[{{Falcon} {et~al.}(2010){Falcon}, {Winget}, {Montgomery}, \&
  {Williams}}]{falcon2010}
{Falcon}, R.~E., {Winget}, D.~E., {Montgomery}, M.~H., \& {Williams}, K.~A.
  2010,
  \href{http://dx.doi.org/10.1088/0004-637X/712/1/585}{\JournalTitle{\apj},
  712, 585}

\bibitem[{{Gaia Collaboration} {et~al.}(2016){Gaia Collaboration}, {Prusti},
  {de Bruijne}, {Brown}, {Vallenari}, {Babusiaux}, {Bailer-Jones}, {Bastian},
  {Biermann}, {Evans}, \& et~al.}]{gaiadr1}
{Gaia Collaboration}, {Prusti}, T., {de Bruijne}, J.~H.~J., {et~al.} 2016,
  \href{http://dx.doi.org/10.1051/0004-6361/201629272}{\JournalTitle{\aap},
  595, A1}

\bibitem[{{Gaia Collaboration} {et~al.}(2018){Gaia Collaboration}, {Brown},
  {Vallenari}, {Prusti}, {de Bruijne}, {Babusiaux}, {Bailer-Jones}, {Biermann},
  {Evans}, {Eyer}, \& et~al.}]{gaiadr2}
{Gaia Collaboration}, {Brown}, A.~G.~A., {Vallenari}, A., {et~al.} 2018,
  \href{http://dx.doi.org/10.1051/0004-6361/201833051}{\JournalTitle{\aap},
  616, A1}

\bibitem[{{Gaia Collaboration} {et~al.}(2021){Gaia Collaboration}, {Brown},
  {Vallenari}, {Prusti}, {de Bruijne}, {Babusiaux}, {Biermann}, {Creevey},
  {Evans}, {Eyer}, {Hutton}, {Jansen}, {Jordi}, {Klioner}, {Lammers},
  {Lindegren}, {Luri}, {Mignard}, {Panem}, {Pourbaix}, {Randich}, {Sartoretti},
  {Soubiran}, {Walton}, {Arenou}, {Bailer-Jones}, {Bastian}, {Cropper},
  {Drimmel}, {Katz}, {Lattanzi}, {van Leeuwen}, {Bakker}, {Cacciari},
  {Casta{\~n}eda}, {De Angeli}, {Ducourant}, {Fabricius}, {Fouesneau},
  {Fr{\'e}mat}, {Guerra}, {Guerrier}, {Guiraud}, {Jean-Antoine Piccolo},
  {Masana}, {Messineo}, {Mowlavi}, {Nicolas}, {Nienartowicz}, {Pailler},
  {Panuzzo}, {Riclet}, {Roux}, {Seabroke}, {Sordo}, {Tanga}, {Th{\'e}venin},
  {Gracia-Abril}, {Portell}, {Teyssier}, {Altmann}, {Andrae}, {Bellas-Velidis},
  {Benson}, {Berthier}, {Blomme}, {Brugaletta}, {Burgess}, {Busso}, {Carry},
  {Cellino}, {Cheek}, {Clementini}, {Damerdji}, {Davidson}, {Delchambre},
  {Dell'Oro}, {Fern{\'a}ndez-Hern{\'a}ndez}, {Galluccio}, {Garc{\'\i}a-Lario},
  {Garcia-Reinaldos}, {Gonz{\'a}lez-N{\'u}{\~n}ez}, {Gosset}, {Haigron},
  {Halbwachs}, {Hambly}, {Harrison}, {Hatzidimitriou}, {Heiter},
  {Hern{\'a}ndez}, {Hestroffer}, {Hodgkin}, {Holl}, {Jan{\ss}en}, {Jevardat de
  Fombelle}, {Jordan}, {Krone-Martins}, {Lanzafame}, {L{\"o}ffler}, {Lorca},
  {Manteiga}, {Marchal}, {Marrese}, {Moitinho}, {Mora}, {Muinonen}, {Osborne},
  {Pancino}, {Pauwels}, {Petit}, {Recio-Blanco}, {Richards}, {Riello},
  {Rimoldini}, {Robin}, {Roegiers}, {Rybizki}, {Sarro}, {Siopis}, {Smith},
  {Sozzetti}, {Ulla}, {Utrilla}, {van Leeuwen}, {van Reeven}, {Abbas}, {Abreu
  Aramburu}, {Accart}, {Aerts}, {Aguado}, {Ajaj}, {Altavilla}, {{\'A}lvarez},
  {{\'A}lvarez Cid-Fuentes}, {Alves}, {Anderson}, {Anglada Varela}, {Antoja},
  {Audard}, {Baines}, {Baker}, {Balaguer-N{\'u}{\~n}ez}, {Balbinot}, {Balog},
  {Barache}, {Barbato}, {Barros}, {Barstow}, {Bartolom{\'e}}, {Bassilana},
  {Bauchet}, {Baudesson-Stella}, {Becciani}, {Bellazzini}, {Bernet}, {Bertone},
  {Bianchi}, {Blanco-Cuaresma}, {Boch}, {Bombrun}, {Bossini}, {Bouquillon},
  {Bragaglia}, {Bramante}, {Breedt}, {Bressan}, {Brouillet}, {Bucciarelli},
  {Burlacu}, {Busonero}, {Butkevich}, {Buzzi}, {Caffau}, {Cancelliere},
  {C{\'a}novas}, {Cantat-Gaudin}, {Carballo}, {Carlucci}, {Carnerero},
  {Carrasco}, {Casamiquela}, {Castellani}, {Castro-Ginard}, {Castro Sampol},
  {Chaoul}, {Charlot}, {Chemin}, {Chiavassa}, {Cioni}, {Comoretto}, {Cooper},
  {Cornez}, {Cowell}, {Crifo}, {Crosta}, {Crowley}, {Dafonte}, {Dapergolas},
  {David}, {David}, {de Laverny}, {De Luise}, {De March}, {De Ridder}, {de
  Souza}, {de Teodoro}, {de Torres}, {del Peloso}, {del Pozo}, {Delbo},
  {Delgado}, {Delgado}, {Delisle}, {Di Matteo}, {Diakite}, {Diener},
  {Distefano}, {Dolding}, {Eappachen}, {Edvardsson}, {Enke}, {Esquej}, {Fabre},
  {Fabrizio}, {Faigler}, {Fedorets}, {Fernique}, {Fienga}, {Figueras},
  {Fouron}, {Fragkoudi}, {Fraile}, {Franke}, {Gai}, {Garabato},
  {Garcia-Gutierrez}, {Garc{\'\i}a-Torres}, {Garofalo}, {Gavras}, {Gerlach},
  {Geyer}, {Giacobbe}, {Gilmore}, {Girona}, {Giuffrida}, {Gomel}, {Gomez},
  {Gonzalez-Santamaria}, {Gonz{\'a}lez-Vidal}, {Granvik},
  {Guti{\'e}rrez-S{\'a}nchez}, {Guy}, {Hauser}, {Haywood}, {Helmi}, {Hidalgo},
  {Hilger}, {H{\l}adczuk}, {Hobbs}, {Holland}, {Huckle}, {Jasniewicz},
  {Jonker}, {Juaristi Campillo}, {Julbe}, {Karbevska}, {Kervella}, {Khanna},
  {Kochoska}, {Kontizas}, {Kordopatis}, {Korn}, {Kostrzewa-Rutkowska},
  {Kruszy{\'n}ska}, {Lambert}, {Lanza}, {Lasne}, {Le Campion}, {Le Fustec},
  {Lebreton}, {Lebzelter}, {Leccia}, {Leclerc}, {Lecoeur-Taibi}, {Liao},
  {Licata}, {Lindstr{\o}m}, {Lister}, {Livanou}, {Lobel}, {Madrero Pardo},
  {Managau}, {Mann}, {Marchant}, {Marconi}, {Marcos Santos}, {Marinoni},
  {Marocco}, {Marshall}, {Martin Polo}, {Mart{\'\i}n-Fleitas}, {Masip},
  {Massari}, {Mastrobuono-Battisti}, {Mazeh}, {McMillan}, {Messina},
  {Michalik}, {Millar}, {Mints}, {Molina}, {Molinaro}, {Moln{\'a}r},
  {Montegriffo}, {Mor}, {Morbidelli}, {Morel}, {Morris}, {Mulone}, {Munoz},
  {Muraveva}, {Murphy}, {Musella}, {Noval}, {Ord{\'e}novic}, {Orr{\`u}},
  {Osinde}, {Pagani}, {Pagano}, {Palaversa}, {Palicio}, {Panahi}, {Pawlak},
  {Pe{\~n}alosa Esteller}, {Penttil{\"a}}, {Piersimoni}, {Pineau}, {Plachy},
  {Plum}, {Poggio}, {Poretti}, {Poujoulet}, {Pr{\v{s}}a}, {Pulone}, {Racero},
  {Ragaini}, {Rainer}, {Raiteri}, {Rambaux}, {Ramos}, {Ramos-Lerate}, {Re
  Fiorentin}, {Regibo}, {Reyl{\'e}}, {Ripepi}, {Riva}, {Rixon}, {Robichon},
  {Robin}, {Roelens}, {Rohrbasser}, {Romero-G{\'o}mez}, {Rowell}, {Royer},
  {Rybicki}, {Sadowski}, {Sagrist{\`a} Sell{\'e}s}, {Sahlmann}, {Salgado},
  {Salguero}, {Samaras}, {Sanchez Gimenez}, {Sanna}, {Santove{\~n}a},
  {Sarasso}, {Schultheis}, {Sciacca}, {Segol}, {Segovia}, {S{\'e}gransan},
  {Semeux}, {Shahaf}, {Siddiqui}, {Siebert}, {Siltala}, {Slezak}, {Smart},
  {Solano}, {Solitro}, {Souami}, {Souchay}, {Spagna}, {Spoto}, {Steele},
  {Steidelm{\"u}ller}, {Stephenson}, {S{\"u}veges}, {Szabados}, {Szegedi-Elek},
  {Taris}, {Tauran}, {Taylor}, {Teixeira}, {Thuillot}, {Tonello}, {Torra},
  {Torra}, {Turon}, {Unger}, {Vaillant}, {van Dillen}, {Vanel}, {Vecchiato},
  {Viala}, {Vicente}, {Voutsinas}, {Weiler}, {Wevers}, {Wyrzykowski}, {Yoldas},
  {Yvard}, {Zhao}, {Zorec}, {Zucker}, {Zurbach}, \& {Zwitter}}]{gaiaedr3}
{Gaia Collaboration}, {Brown}, A.~G.~A., {Vallenari}, A., {et~al.} 2021,
  \href{http://dx.doi.org/10.1051/0004-6361/202039657}{\JournalTitle{\aap},
  649, A1}

\bibitem[{{Genest-Beaulieu} \& {Bergeron}(2019)}]{gb2019}
{Genest-Beaulieu}, C. \& {Bergeron}, P. 2019,
  \href{http://dx.doi.org/10.3847/1538-4357f/aafac6}{\JournalTitle{\apj}, 871,
  169}

\bibitem[{{Gentile Fusillo} {et~al.}(2019){Gentile Fusillo}, {Tremblay},
  {G{\"a}nsicke}, {Manser}, {Cunningham}, {Cukanovaite}, {Hollands}, {Marsh},
  {Raddi}, {Jordan}, {Toonen}, {Geier}, {Barstow}, \& {Cummings}}]{gf2019}
{Gentile Fusillo}, N.~P., {Tremblay}, P.-E., {G{\"a}nsicke}, B.~T., {et~al.}
  2019, \href{http://dx.doi.org/10.1093/mnras/sty3016}{\JournalTitle{\mnras},
  482, 4570}

\bibitem[{{Godoy-Rivera} \& {Chanam{\'e}}(2018)}]{gr2018}
{Godoy-Rivera}, D. \& {Chanam{\'e}}, J. 2018,
  \href{http://dx.doi.org/10.1093/mnras/sty1736}{\JournalTitle{\mnras}, 479,
  4440}

\bibitem[{{Godoy-Rivera} {et~al.}(2021){Godoy-Rivera}, {Tayar}, {Pinsonneault},
  {Rodr{\'\i}guez Mart{\'\i}nez}, {Stassun}, {van Saders}, {Beaton},
  {Garc{\'\i}a-Hern{\'a}ndez}, \& {Teske}}]{gr2021}
{Godoy-Rivera}, D., {Tayar}, J., {Pinsonneault}, M.~H., {et~al.} 2021,
  \href{http://dx.doi.org/10.3847/1538-4357/abf8ba}{\JournalTitle{\apj}, 915,
  19}

\bibitem[{{Greenstein}(1986)}]{greenstein1986}
{Greenstein}, J.~L. 1986,
  \href{http://dx.doi.org/10.1086/114219}{\JournalTitle{\aj}, 92, 859}

\bibitem[{{Gustafsson} {et~al.}(2008){Gustafsson}, {Edvardsson}, {Eriksson},
  {J{\o}rgensen}, {Nordlund}, \& {Plez}}]{gustafson2008}
{Gustafsson}, B., {Edvardsson}, B., {Eriksson}, K., {et~al.} 2008,
  \href{http://dx.doi.org/10.1051/0004-6361:200809724}{\JournalTitle{\aap},
  486, 951}

\bibitem[{{Halenka} {et~al.}(2015){Halenka}, {Olchawa}, {Madej}, \&
  {Grabowski}}]{halenka2015}
{Halenka}, J., {Olchawa}, W., {Madej}, J., \& {Grabowski}, B. 2015,
  \href{http://dx.doi.org/10.1088/0004-637X/808/2/131}{\JournalTitle{\apj},
  808, 131}

\bibitem[{{Jeffries} \& {Stevens}(1996)}]{js1996}
{Jeffries}, R.~D. \& {Stevens}, I.~R. 1996,
  \href{http://dx.doi.org/10.1093/mnras/279.1.180}{\JournalTitle{\mnras}, 279,
  180}

\bibitem[{{Jiang} \& {Tremaine}(2010)}]{jt2010}
{Jiang}, Y.-F. \& {Tremaine}, S. 2010,
  \href{http://dx.doi.org/10.1111/j.1365-2966.2009.15744.x}{\JournalTitle{\mnras},
  401, 977}

\bibitem[{{J{\o}rgensen} \& {Lindegren}(2005)}]{jl2005}
{J{\o}rgensen}, B.~R. \& {Lindegren}, L. 2005,
  \href{http://dx.doi.org/10.1051/0004-6361:20042185}{\JournalTitle{\aap}, 436,
  127}

\bibitem[{{Kalirai} {et~al.}(2007){Kalirai}, {Bergeron}, {Hansen}, {Kelson},
  {Reitzel}, {Rich}, \& {Richer}}]{kalirai2007}
{Kalirai}, J.~S., {Bergeron}, P., {Hansen}, B. M.~S., {et~al.} 2007,
  \href{http://dx.doi.org/10.1086/521922}{\JournalTitle{\apj}, 671, 748}

\bibitem[{{Kalirai} {et~al.}(2005){Kalirai}, {Richer}, {Reitzel}, {Hansen},
  {Rich}, {Fahlman}, {Gibson}, \& {von Hippel}}]{kalirai2005}
{Kalirai}, J.~S., {Richer}, H.~B., {Reitzel}, D., {et~al.} 2005,
  \href{http://dx.doi.org/10.1086/427774}{\JournalTitle{\apjl}, 618, L123}

\bibitem[{{Kepler} {et~al.}(2015){Kepler}, {Pelisoli}, {Koester}, {Ourique},
  {Kleinman}, {Romero}, {Nitta}, {Eisenstein}, {Costa}, {K{\"u}lebi}, {Jordan},
  {Dufour}, {Giommi}, \& {Rebassa-Mansergas}}]{kepler2015}
{Kepler}, S.~O., {Pelisoli}, I., {Koester}, D., {et~al.} 2015,
  \href{http://dx.doi.org/10.1093/mnras/stu2388}{\JournalTitle{\mnras}, 446,
  4078}

\bibitem[{{Kepler} {et~al.}(2016){Kepler}, {Pelisoli}, {Koester}, {Ourique},
  {Romero}, {Reindl}, {Kleinman}, {Eisenstein}, {Valois}, \&
  {Amaral}}]{kepler2016}
{Kepler}, S.~O., {Pelisoli}, I., {Koester}, D., {et~al.} 2016,
  \href{http://dx.doi.org/10.1093/mnras/stv2526}{\JournalTitle{\mnras}, 455,
  3413}

\bibitem[{{Kilic} {et~al.}(2021{\natexlab{a}}){Kilic}, {B{\'e}dard}, \&
  {Bergeron}}]{kilic2021a}
{Kilic}, M., {B{\'e}dard}, A., \& {Bergeron}, P. 2021{\natexlab{a}},
  \href{http://dx.doi.org/10.1093/mnras/stab439}{\JournalTitle{\mnras}, 502,
  4972}

\bibitem[{{Kilic} {et~al.}(2021{\natexlab{b}}){Kilic}, {Bergeron}, {Blouin}, \&
  {B{\'e}dard}}]{kilic2021b}
{Kilic}, M., {Bergeron}, P., {Blouin}, S., \& {B{\'e}dard}, A.
  2021{\natexlab{b}},
  \href{http://dx.doi.org/10.1093/mnras/stab767}{\JournalTitle{\mnras}, 503,
  5397}

\bibitem[{{Kilic} {et~al.}(2019){Kilic}, {Bergeron}, {Dame}, {Hambly},
  {Rowell}, \& {Crawford}}]{kilic2019}
{Kilic}, M., {Bergeron}, P., {Dame}, K., {et~al.} 2019,
  \href{http://dx.doi.org/10.1093/mnras/sty2755}{\JournalTitle{\mnras}, 482,
  965}

\bibitem[{{Kilic} {et~al.}(2018){Kilic}, {Hambly}, {Bergeron},
  {Genest-Beaulieu}, \& {Rowell}}]{kilic2018}
{Kilic}, M., {Hambly}, N.~C., {Bergeron}, P., {Genest-Beaulieu}, C., \&
  {Rowell}, N. 2018,
  \href{http://dx.doi.org/10.1093/mnrasl/sly110}{\JournalTitle{\mnras}, 479,
  L113}

\bibitem[{{Kleinman} {et~al.}(2004){Kleinman}, {Harris}, {Eisenstein},
  {Liebert}, {Nitta}, {Krzesi{\'n}ski}, {Munn}, {Dahn}, {Hawley}, {Pier},
  {Schmidt}, {Silvestri}, {Smith}, {Szkody}, {Strauss}, {Knapp}, {Collinge},
  {Mukadam}, {Koester}, {Uomoto}, {Schlegel}, {Anderson}, {Brinkmann}, {Lamb},
  {Schneider}, \& {York}}]{kleinman2004}
{Kleinman}, S.~J., {Harris}, H.~C., {Eisenstein}, D.~J., {et~al.} 2004,
  \href{http://dx.doi.org/10.1086/383464}{\JournalTitle{\apj}, 607, 426}

\bibitem[{{Kleinman} {et~al.}(2013){Kleinman}, {Kepler}, {Koester}, {Pelisoli},
  {Pe{\c c}anha}, {Nitta}, {Costa}, {Krzesinski}, {Dufour}, {Lachapelle},
  {Bergeron}, {Yip}, {Harris}, {Eisenstein}, {Althaus}, \&
  {C{\'o}rsico}}]{kleinman2013}
{Kleinman}, S.~J., {Kepler}, S.~O., {Koester}, D., {et~al.} 2013,
  \href{http://dx.doi.org/10.1088/0067-0049/204/1/5}{\JournalTitle{\apjs}, 204,
  5}

\bibitem[{{Koester}(2010)}]{koester2010}
{Koester}, D. 2010, \JournalTitle{\memsai}, 81, 921

\bibitem[{{Kraus} \& {Hillenbrand}(2009)}]{kh2009}
{Kraus}, A.~L. \& {Hillenbrand}, L.~A. 2009,
  \href{http://dx.doi.org/10.1088/0004-637X/704/1/531}{\JournalTitle{\apj},
  704, 531}

\bibitem[{{Kunder} {et~al.}(2017){Kunder}, {Kordopatis}, {Steinmetz},
  {Zwitter}, {McMillan}, {Casagrande}, {Enke}, {Wojno}, {Valentini},
  {Chiappini}, {Matijevi{\v{c}}}, {Siviero}, {de Laverny}, {Recio-Blanco},
  {Bijaoui}, {Wyse}, {Binney}, {Grebel}, {Helmi}, {Jofre}, {Antoja}, {Gilmore},
  {Siebert}, {Famaey}, {Bienaym{\'e}}, {Gibson}, {Freeman}, {Navarro},
  {Munari}, {Seabroke}, {Anguiano}, {{\v{Z}}erjal}, {Minchev}, {Reid},
  {Bland-Hawthorn}, {Kos}, {Sharma}, {Watson}, {Parker}, {Scholz}, {Burton},
  {Cass}, {Hartley}, {Fiegert}, {Stupar}, {Ritter}, {Hawkins}, {Gerhard},
  {Chaplin}, {Davies}, {Elsworth}, {Lund}, {Miglio}, \& {Mosser}}]{rave2017}
{Kunder}, A., {Kordopatis}, G., {Steinmetz}, M., {et~al.} 2017,
  \href{http://dx.doi.org/10.3847/1538-3881/153/2/75}{\JournalTitle{\aj}, 153,
  75}

\bibitem[{{Lachaume} {et~al.}(1999){Lachaume}, {Dominik}, {Lanz}, \&
  {Habing}}]{lachaume1999}
{Lachaume}, R., {Dominik}, C., {Lanz}, T., \& {Habing}, H.~J. 1999,
  \JournalTitle{\aap}, 348, 897

\bibitem[{{Li} {et~al.}(2020){Li}, {Bedding}, {Christensen-Dalsgaard},
  {Stello}, {Li}, \& {Keen}}]{tandali2020}
{Li}, T., {Bedding}, T.~R., {Christensen-Dalsgaard}, J., {et~al.} 2020,
  \href{http://dx.doi.org/10.1093/mnras/staa1350}{\JournalTitle{\mnras}, 495,
  3431}

\bibitem[{{Liebert} {et~al.}(2005){Liebert}, {Bergeron}, \&
  {Holberg}}]{liebert2005}
{Liebert}, J., {Bergeron}, P., \& {Holberg}, J.~B. 2005,
  \href{http://dx.doi.org/10.1086/425738}{\JournalTitle{\apjs}, 156, 47}

\bibitem[{{Luo} {et~al.}(2019){Luo}, {Zhao}, {Zhao}, \& {et al.}}]{lamost2019}
{Luo}, A.~L., {Zhao}, Y.~H., {Zhao}, G., \& {et al.} 2019, \JournalTitle{VizieR
  Online Data Catalog}, V/164

\bibitem[{{Marigo} {et~al.}(2020){Marigo}, {Cummings}, {Curtis}, {Kalirai},
  {Chen}, {Tremblay}, {Ramirez-Ruiz}, {Bergeron}, {Bladh}, {Bressan},
  {Girardi}, {Pastorelli}, {Trabucchi}, {Cheng}, {Aringer}, \&
  {Tio}}]{marigo2020}
{Marigo}, P., {Cummings}, J.~D., {Curtis}, J.~L., {et~al.} 2020,
  \href{http://dx.doi.org/10.1038/s41550-020-1132-1}{\JournalTitle{Nature
  Astronomy}, 4, 1102}

\bibitem[{Mathur {et~al.}(2017)Mathur, Huber, Batalha, Ciardi, Bastien,
  Bieryla, Buchhave, Cochran, Endl, Esquerdo, Furlan, Howard, Howell, Isaacson,
  Latham, MacQueen, \& Silva}]{Mathur2017}
Mathur, S., Huber, D., Batalha, N.~M., {et~al.} 2017,
  \href{http://dx.doi.org/10.3847/1538-4365/229/2/30}{\JournalTitle{The
  Astrophysical Journal Supplement Series}, 229, 30}

\bibitem[{{Michalik} {et~al.}(2015){Michalik}, {Lindegren}, \&
  {Hobbs}}]{tgas2015}
{Michalik}, D., {Lindegren}, L., \& {Hobbs}, D. 2015,
  \href{http://dx.doi.org/10.1051/0004-6361/201425310}{\JournalTitle{\aap},
  574, A115}

\bibitem[{{Munn} {et~al.}(2014){Munn}, {Harris}, {von Hippel}, {Kilic},
  {Liebert}, {Williams}, {DeGenarro}, {Jeffery}, \& {Tilleman}}]{munn2014}
{Munn}, J.~A., {Harris}, H.~C., {von Hippel}, T., {et~al.} 2014,
  \href{http://dx.doi.org/10.1088/0004-6256/148/6/132}{\JournalTitle{\aj}, 148,
  132}

\bibitem[{{Napiwotzki} {et~al.}(2020){Napiwotzki}, {Karl}, {Lisker},
  {Catal{\'a}n}, {Drechsel}, {Heber}, {Homeier}, {Koester}, {Leibundgut},
  {Marsh}, {Moehler}, {Nelemans}, {Reimers}, {Renzini}, {Str{\"o}er}, \&
  {Yungelson}}]{napi2020}
{Napiwotzki}, R., {Karl}, C.~A., {Lisker}, T., {et~al.} 2020,
  \href{http://dx.doi.org/10.1051/0004-6361/201629648}{\JournalTitle{\aap},
  638, A131}

\bibitem[{{Navarrete} {et~al.}(2015){Navarrete}, {Chanam{\'e}}, {Ram{\'\i}rez},
  {Meza}, {Anglada-Escud{\'e}}, \& {Shkolnik}}]{navarrete2015}
{Navarrete}, C., {Chanam{\'e}}, J., {Ram{\'\i}rez}, I., {et~al.} 2015,
  \href{http://dx.doi.org/10.1088/0004-637X/808/1/103}{\JournalTitle{\apj},
  808, 103}

\bibitem[{{Nordstr{\"o}m} {et~al.}(2004){Nordstr{\"o}m}, {Mayor}, {Andersen},
  {Holmberg}, {Pont}, {J{\o}rgensen}, {Olsen}, {Udry}, \&
  {Mowlavi}}]{nordstrom2004}
{Nordstr{\"o}m}, B., {Mayor}, M., {Andersen}, J., {et~al.} 2004,
  \href{http://dx.doi.org/10.1051/0004-6361:20035959}{\JournalTitle{\aap}, 418,
  989}

\bibitem[{{Quinn} {et~al.}(2010){Quinn}, {Wilkinson}, {Irwin}, {Marshall},
  {Koch}, \& {Belokurov}}]{quinn2010}
{Quinn}, D.~P., {Wilkinson}, M.~I., {Irwin}, M.~J., {et~al.} 2010, in
  Astronomical Society of the Pacific Conference Series, Vol. 435, Binaries -
  Key to Comprehension of the Universe, ed. A.~{Pr{\v{s}}a} \& M.~{Zejda}, 453

\bibitem[{{Ram{\'\i}rez} {et~al.}(2019){Ram{\'\i}rez}, {Khanal}, {Lichon},
  {Chanam{\'e}}, {Endl}, {Mel{\'e}ndez}, \& {Lambert}}]{ramirez2019}
{Ram{\'\i}rez}, I., {Khanal}, S., {Lichon}, S.~J., {et~al.} 2019,
  \href{http://dx.doi.org/10.1093/mnras/stz2709}{\JournalTitle{\mnras}, 490,
  2448}

\bibitem[{{Ram{\'{\i}}rez} {et~al.}(2014){Ram{\'{\i}}rez}, {Mel{\'e}ndez},
  {Bean}, {Asplund}, {Bedell}, {Monroe}, {Casagrande}, {Schirbel}, {Dreizler},
  {Teske}, {Tucci Maia}, {Alves-Brito}, \& {Baumann}}]{ramirez2014}
{Ram{\'{\i}}rez}, I., {Mel{\'e}ndez}, J., {Bean}, J., {et~al.} 2014,
  \href{http://dx.doi.org/10.1051/0004-6361/201424244}{\JournalTitle{\aap},
  572, A48}

\bibitem[{{Renedo} {et~al.}(2010){Renedo}, {Althaus}, {Miller Bertolami},
  {Romero}, {C{\'o}rsico}, {Rohrmann}, \& {Garc{\'{\i}}a-Berro}}]{renedo2010}
{Renedo}, I., {Althaus}, L.~G., {Miller Bertolami}, M.~M., {et~al.} 2010,
  \href{http://dx.doi.org/10.1088/0004-637X/717/1/183}{\JournalTitle{\apj},
  717, 183}

\bibitem[{{Romero} {et~al.}(2019){Romero}, {Kepler}, {Joyce}, {Lauffer}, \&
  {C{\'o}rsico}}]{romero2019}
{Romero}, A.~D., {Kepler}, S.~O., {Joyce}, S.~R.~G., {Lauffer}, G.~R., \&
  {C{\'o}rsico}, A.~H. 2019,
  \href{http://dx.doi.org/10.1093/mnras/stz160}{\JournalTitle{\mnras}, 484,
  2711}

\bibitem[{{Silvestri} {et~al.}(2001){Silvestri}, {Oswalt}, {Wood}, {Smith},
  {Reid}, \& {Sion}}]{silvestri2001}
{Silvestri}, N.~M., {Oswalt}, T.~D., {Wood}, M.~A., {et~al.} 2001,
  \href{http://dx.doi.org/10.1086/318005}{\JournalTitle{\aj}, 121, 503}

\bibitem[{{Sneden}(1973)}]{sneden1973}
{Sneden}, C.~A. 1973, PhD thesis, THE UNIVERSITY OF TEXAS AT AUSTIN.

\bibitem[{{Soubiran} {et~al.}(2016){Soubiran}, {Le Campion}, {Brouillet}, \&
  {Chemin}}]{soubiran2016}
{Soubiran}, C., {Le Campion}, J.-F., {Brouillet}, N., \& {Chemin}, L. 2016,
  \href{http://dx.doi.org/10.1051/0004-6361/201628497}{\JournalTitle{\aap},
  591, A118}

\bibitem[{{Tremblay} {et~al.}(2013){Tremblay}, {Ludwig}, {Steffen}, \&
  {Freytag}}]{tremblay2013}
{Tremblay}, P.~E., {Ludwig}, H.~G., {Steffen}, M., \& {Freytag}, B. 2013,
  \href{http://dx.doi.org/10.1051/0004-6361/201322318}{\JournalTitle{\aap},
  559, A104}

\bibitem[{{Weidemann}(2000)}]{weidemann2000}
{Weidemann}, V. 2000, \JournalTitle{\aap}, 363, 647

\bibitem[{{Weiss} \& {Ferguson}(2009)}]{weiss2009}
{Weiss}, A. \& {Ferguson}, J.~W. 2009,
  \href{http://dx.doi.org/10.1051/0004-6361/200912043}{\JournalTitle{\aap},
  508, 1343}

\bibitem[{{Williams} {et~al.}(2009){Williams}, {Bolte}, \&
  {Koester}}]{williams2009}
{Williams}, K.~A., {Bolte}, M., \& {Koester}, D. 2009,
  \href{http://dx.doi.org/10.1088/0004-637X/693/1/355}{\JournalTitle{\apj},
  693, 355}

\bibitem[{{Williams} {et~al.}(2018){Williams}, {Canton}, {Bellini}, {Bolte},
  {Rubin}, {Gianninas}, \& {Kilic}}]{williams2018}
{Williams}, K.~A., {Canton}, P.~A., {Bellini}, A., {et~al.} 2018,
  \href{http://dx.doi.org/10.3847/1538-4357/aad90b}{\JournalTitle{\apj}, 867,
  62}

\bibitem[{{Yi} {et~al.}(2001){Yi}, {Demarque}, {Kim}, {Lee}, {Ree}, {Lejeune},
  \& {Barnes}}]{yi2001}
{Yi}, S., {Demarque}, P., {Kim}, Y.-C., {et~al.} 2001,
  \href{http://dx.doi.org/10.1086/321795}{\JournalTitle{\apjs}, 136, 417}

\bibitem[{{Yoo} {et~al.}(2004){Yoo}, {Chanam{\'e}}, \& {Gould}}]{YCG2004}
{Yoo}, J., {Chanam{\'e}}, J., \& {Gould}, A. 2004,
  \href{http://dx.doi.org/10.1086/380562}{\JournalTitle{\apj}, 601, 311}

\bibitem[{{Zhao} {et~al.}(2012){Zhao}, {Oswalt}, {Willson}, {Wang}, \&
  {Zhao}}]{zhao2012}
{Zhao}, J.~K., {Oswalt}, T.~D., {Willson}, L.~A., {Wang}, Q., \& {Zhao}, G.
  2012,
  \href{http://dx.doi.org/10.1088/0004-637X/746/2/144}{\JournalTitle{\apj},
  746, 144}

\end{thebibliography}




\end{document}